\documentclass[AMS]{mn2e}
\usepackage{graphicx}

\def\lesssim{\buildrel < \over {_{\sim}}}
\def\gtrsim{\buildrel > \over {_{\sim}}}

\begin{document}

\title[Dark stars in the dark ages]{Dark matter annihilation effects on the first stars}

\author[]{F. Iocco$^{1}$, A. Bressan$^{2,3,4}$, E. Ripamonti$^{5}$, R. Schneider$^{1}$, A. Ferrara$^{4}$, P. Marigo$^{6}$\\
$^{1}$INAF/Osservatorio Astrofisico di Arcetri; Largo Enrico Fermi 5, Firenze, Italy\\
$^{2}$INAF/Osservatorio Astronomico di Padova; Vicolo dell'Osservatorio 5, Padova, Italy\\
$^{3}$INAOE; Luis Enrique Erro 1, 72840, Tonantzintla, Puebla, Mexico\\
$^{4}$SISSA; Via Beirut 4, Trieste, Italy\\
$^{5}$Universit\`a degli Studi dell'Insubria, Dip. di Scienze Chimiche, Fisiche e Naturali; Via Valleggio 12, Como, Italy\\
$^{6}$Universit\`a degli Studi di Padova, Dip. di Astronomia; Vicolo dell'Osservatorio 3, Padova, Italy}

\pagerange{\pageref{firstpage}--\pageref{lastpage}} \pubyear{2008}
\maketitle


\begin{abstract}
We study the effects of weakly interacting massive particles (WIMP) dark matter (DM) on the collapse and evolution of the first stars in the Universe. 
Using a stellar evolution code, we follow the pre-Main Sequence (MS) phase of a grid of 
metal--free stars with masses in the range $5 M_{\odot} \leq M_{\ast} \leq 600 M_{\odot}$ forming in the
centre of a $10^6 M_{\odot}$ halo at $z = 20$.  DM particles of the parent halo are accreted in the 
proto-stellar interior by adiabatic contraction and scattering/capture processes, reaching central 
densities of ${\cal O}(10^{12}$ GeV cm$^{-3}$) at radii of the order of 10 AU. 
Energy release from annihilation reactions can effectively 
counteract the gravitational collapse, in agreement with results from other groups. 
We find this stalling phase (known as a {\it dark} star) is transient
and lasts from $2.1 \times 10^3$yr ($M_{\ast}=600 M_\odot$) to $1.8 \times 10^4$yr ($M_{\ast}=9 M_\odot$). 
Later in the evolution, DM scattering/capture rate becomes high enough that energy deposition from annihilations 
significantly alters the pre-MS evolution of the star in a way that depends on DM (i) velocity dispersion, 
$\bar v$, (ii) density, $\rho$, (iii) elastic scattering cross section with baryons, $\sigma_0$. 
For our fiducial set of parameters ($\bar v, \rho, \sigma_0$)= ($10$ km s$^{-1}$, $10^{11}$ GeV cm$^{-3}$, $10^{-38}$ cm$^2$) 
we find that the evolution of stars of mass $ M_{\ast} < 40 M_\odot$ ``freezes'' on the HR diagram before reaching the ZAMS. 
Stars with  $M_{\ast} \geq 40 M_\odot$ manage to ignite nuclear reactions; however, DM ``burning'' prolonges their lifetimes by a 
factor 2 (5) for a $600 M_{\odot}$ ($40 M_{\odot}$) star.
\end{abstract}


\section{Introduction}
Current observations of primordial light element abundances, baryon acoustic oscillations, distance 
measurements by means of Type Ia supernovae and cosmic microwave background, all fit together to describe 
a Universe undergoing an accelerated expansion (see Spergel et al. 2007; Komatsu et al. 2008 and references 
therein). 
An unknown energy field, often referred to as Dark Energy, constitutes approximately 75\% of the total 
energy density, whereas the remaining is made of matter. However, only 15\% of the latter is made of 
known particles (baryons): its majority seems to be composed of a non-visible, unknown component, commonly 
referred to as Dark Matter (DM). If thermally produced in the hot plasma, models best fitting observations 
require it to have decoupled at temperatures much smaller than its mass, thus being often referred to as 
{\it cold}. 

In this scenario, small scale perturbations grow faster and detach first from the Hubble flow, leading to 
a ``hierarchical'' growth of structures starting off very small haloes in the young universe, and building up 
bigger ones by means of mergers. The first stars are predicted to form at $z < 20 - 30$ in
haloes with masses $M = 10^6 - 10^8 M_{\odot}$, generally referred to as mini--haloes (see Barkana \& Loeb 2001
and Ciardi \& Ferrara 2005 for thorough reviews of the subject). The gas virialized in the potential wells
of these systems has primordial chemical composition and low temperatures, $T_{\rm vir} < 10^4$~K; in these conditions
the additional cooling necessary for the gas to collapse and form stars is provided by molecular hydrogen.
The results of recent semi-analytic studies and of sophisticated 3D numerical simulations consistently indicate 
that the reduced cooling efficiency, together with the absence of magnetic fields and of relevant angular 
momentum effects, inhibits gas fragmentation and lead to the formation of a single massive star
(Omukai \& Nishi 1998; Bromm, Coppi \& Larson 1999, 2001; Abel et al. 2000, 2002; Nakamura \& Umemura 2001;
Ripamonti et al. 2002; Gao et al. 2007; O'Shea \& Norman 2007; Yoshida et al. 2006, 2007). 
The mass of these first stars, often called Population III (Pop III) stars, is still uncertain but likely 
to be in the range $30 M_{\odot} < M_{\ast} < 300 M_{\odot}$, depending on the strength of feedback effects 
which control the mass growth by accretion on the central high-density core (Omukai \& Palla 2003; 
Tan \& McKee 2004).      

Since Pop III stars are predicted to form at high $z$ or are hidden in the outskirts
of collapsing structures at moderate $z$ (Schneider et al. 2006; Tornatore, Schneider \& Ferrara 2007), 
observational evidences of their nature and properties are still lacking.
If very massive, they are thought to explode as powerful pair--instability supernovae 
or to directly collapse to black holes after a short lifetime of a few Myr \cite{Heger:2001cd}. 
The chemical imprint they leave on subsequent stellar generations is difficult to identify
in current samples of extremely metal--poor stars in the Galactic halo 
(Tumlinson 2006; Salvadori, Schneider \& Ferrara 2007); their signature on the 
reionization history is weak (Gnedin \& Fan 2006; Choudhury \& Ferrara 2006), and so is any feature 
in the low an high energy diffuse neutrino background (Schneider, Guetta \& Ferrara 2001; 
Iocco et al. 2005, 2008). 
Future probes of the nature of Pop III stars will come from the James Webb Space Telescope 
or from 21cm telescopes which are expected to operate within the next decade.

Intricate as the ``standard'' scenario can be, with ordinary matter only gravitationally interacting 
with its dark counterpart, there are instances where the situation could have been complicated by 
additional interactions between dark and ordinary matter.

There is a flourishing zoology of models providing candidates for DM particles, and we address the reader 
to Bertone, Hooper \& Silk (2005) for a thorough review of motivations, candidates, and their properties. The 
currently favored model, which naturally complies with the requirements arising from cosmological and 
particle physics arguments, is the lightest stable particle in a supersymmetric extension of the 
standard model of particle physics.
Often referred to as \emph{neutralinos}, these are Majorana particles coupled to baryons by means of weak 
interactions, with the most remarkable properties to be self--annihilating, and to have a non--vanishing 
scattering cross--section with standard model particles. 

For what concerns this paper, the consequence of these properties is two--fold: 
in environments with high enough density of DM particles, self--annihilation of neutralinos 
could constitute a source of energy, emitted in the form of radiation, which can potentially overcome 
the cooling of the gas, inhibiting or slowing down the formation of a proto--star. 
When (and if) something resembling a celestial object is formed, such weakly interacting massive particles 
(WIMPs) can scatter off the baryonic material and lose energy, thus being gravitationally captured; 
they accumulate and annihilate inside the object, thus providing it with an (additional) energy source.

At present, the typical DM densities are too low to provide any dramatic, widespread 
effect on stellar evolution: recent calculations show that DM densities necessary to induce 
strong effects are achievable only within the central two parsecs of our galaxy, with similar restrictions 
applying to all galaxies in the Local Universe (Fairbairn, Scott \& Edsjo, 2008). 
The first star--forming mini-haloes, however, are smaller and denser: this provides favorable conditions 
for DM annihilation effects to play a role. 

As noticed by Spolyar, Freese \& Gondolo (2008), 
WIMP annihilation in young haloes during the formation of the first stars could provide an amount of 
energy equal to the one dissipated by chemical cooling of the gas. Also, as noticed by Iocco (2008) and 
Freese, Spolyar \& Aguirre (2008), the energy produced by DM 
annihilation captured inside early stars could even exceed the one produced by their nuclear burning.

Additional effects of DM decays and annihilations have also been considered in the literature, 
with particular attention to their contribution on the reionization and thermal histories of the intergalactic 
medium (Mapelli \& Ferrara 2005; Ripamonti, Mapelli \& Ferrara 2007a; Vald\'es et al. 2007) as well 
as on the conditions for star formation in the first mini--haloes (Ripamonti, Mapelli \& Ferrara 2007b).  

In this paper, we aim at studying the effects of neutralino DM annihilation on stellar 
evolution in the early universe. We will treat separately the effects arising from DM 
contraction during stellar collapse and from DM captured by scattering with the baryons: 
although these are clearly part of the same picture, the physical mechanisms are different and 
observational and experimental constraints on the parameters involved have different nature and 
reliability. 

The paper is organized as follows: in Sec. \ref{inimodel} we describe the initial conditions
of the model; in Sec. \ref{adiabatic} we introduce the process of adiabatic contraction and
discuss the evolution of our fiducial $100 M_{\odot}$ star in the presence of this mechanism. 
Sec. \ref{capture} deals with the scattering/capture process and its impact on the evolution
of the fiducial stellar model. In Sec. \ref{evolution} we discuss the dependences on the
stellar mass and DM parameters; in Sec. \ref{DMconstrSF} we summarize 
the effects of scattering/capture process on stellar models. 
Finally, in Sec. \ref{Concl} we discuss our conclusions.
We defer to the Appendixes a synthetic description of the stellar evolutionary code and 
include tables with relevant physical quantities of all stellar models considered in this
study.

Throughout the paper we work in the framework of a $\Lambda$ cold DM ($\Lambda$CDM) 
cosmological model with parameters $\Omega_M = 0.24$, $\Omega_{\Lambda}=0.76$, $\Omega_B=0.042$,
$h=0.73$ (Spergel et al. 2007; Hinshaw et al. 2008) and we assume that DM is entirely made of 
neutralinos with a mass of 100 GeV.

\section{Initial conditions}
\label{inimodel}

In this Section we describe our initial conditions for the density profile of the dark matter halo 
and for the pre--Main Sequence (pre--MS) evolution of the star. In the rest of the paper we often refer
to the object under investigation as a star, or dark star, although, in most cases, it is 
actually a proto--star. We will make the distinction clear where necessary.

\subsection{Dark matter halo}
\label{haloinimod}

In the present study, we implement the 
characteristics of a ``standard'' early star forming mini--halo,  
i.e. an object with total mass $M =10^6 M_{\odot}$ virializing at $z=20$ 
(see e.g. Abel et al. 2002; Bromm et al. 2002; Yoshida et al. 2006; 
Gao et al. 2007; Turk 2007). We assume that 82.5\% of the total mass 
is DM, while the rest is baryonic; DM follows a standard 
NFW profile (Navarro, Frenk \& White 1996), with virial radius 
$R_{\rm vir} = R_{200} = 5 \times 10^{20}$~cm and concentration\footnote{We note that the ``adiabatically contracted'' DM profiles are 
almost independent of $c$ (at least if $c\lesssim100-1000$).} $c=10$.
We flatten this profile for radii smaller than the free--streaming length 
of DM particles ($6.6 \times 10^{12}$~cm if their mass is 100~GeV). 

It is worth noting that approximately $100 M_\odot$ of dark matter, 
equivalent to the mass of the fiducial stellar model that we will
discuss in the following Sections, are contained within a radius of $\approx 10^{18}$~cm. 
This qualitatively defines the maximum distance at which DM particles feel 
the gravitational pull due to the central concentration of baryons or, 
in other words, the maximum radius where adiabatic contraction effects are 
relevant, as described in Sec. \ref{adiabatic}.

\subsection{Proto-star}
\label{inistemod}

We assume that in the first mini-haloes stars form as a result of the collapse of metal--free gas 
clouds, after the cooling--induced fragmentation phase is completed (see e.g. Omukai 2000 and 
Schneider et al. 2002).
We adopt the Padova Stellar Evolution code in the version suitable for the study of zero metallicity stars 
(Marigo et al. 2001, 2003). A synthetic description of the code with details on the computed evolutionary 
tracks is presented in Appendix \ref{StelCoDet}.

In order to catch the proto--star as early as possible in its evolution, 
within the convergence limits of the code and its physical reliability,  for each stellar model we 
force its thermodynamic conditions to the tip of the Hayashi track by
providing a density-dependent heating source; this causes an expansion of 
the proto-star and a drop of the effective temperature. 
We stress that this initial phase has no particular physical meaning:
any other convenient artificial heating source would work to this purpose.

We prepare the initial conditions as follows. Starting from the configuration of a Zero Age Main Sequence 
(ZAMS) star of the same mass and adopting a primordial chemical composition\footnote{We assume a H and He 
mass fractions of $X=0.755$ and $Y=0.245$, respectively, according to recent BBN models \cite{Iocco:2007km}.} 
we then artificially expand the star towards the tip of the Hayashi track.  
During this artificial evolution we perform a preliminary check on the relative strengths of the DM and 
gravitational luminosities that, during this phase, correspond (but not exactly) to the stellar luminosity. 
For a $100 M_\odot$ star, which we consider our fiducial model, the DM luminosity decreases as the 
baryonic configuration gets more expanded.
We continue the artificial expansion until  the ratio of DM annihilation 
to the total stellar one is $L_{\rm DM}/L_\ast \leq 0.5$. At this stage, the central 
temperature is $T_c \approx 5 \times 10^4$~K and the central gas density 
is $\rho_c \approx 10^{-7}$~g cm$^{-3}$. 
The radius of the object is $\approx 10^{14}$~cm and, according to the DM profile 
described in the previous Section, the enclosed DM mass is $\approx 10^{31}$~g, only 10$^{-4}$ of the
proto--stellar  mass.
It is difficult to push the proto--star beyond this point because of numerical problems.

Starting from this configuration, we follow the pre--MS contraction phase, including the DM annihilation energy 
source term in the structure equations, as described in Sections \ref{adiabatic} and \ref{capture}. 
It is important to stress that this model is physically self--consistent: if gravitational energy 
release is the only luminosity source, the model correctly reproduces the usual track of a 
``standard", non DM--burning proto--star in the HR diagram.

\section{Dark matter contraction}
\label{adiabatic}

As we have discussed in the previous Section, in a proto--star located at the center of the halo the DM content 
is $\approx$10$^{-4}$ of the total mass; as a result, the DM contribution to the gravitational potential is 
negligible, as this is largely dominated by the baryonic mass.
Starting from the initial profile described in Sec. \ref{haloinimod}, we evolve the density profile 
using the so-called adiabatic contraction (AC) approximation (Blumenthal et al. 1986), 
which is based on the assumption that the orbital time of the particles is much longer than the infall 
time (namely, orbits never cross each other). In this Section, we first introduce the formalism adopted
to implement AC in our model and then describe the evolution of our fiducial $100 M_{\odot}$ proto-star
in the presence of DM contraction.

\subsection{Formalism and approximations} 

The AC approximation identifies the adiabatic invariant $M(R)R$, where $M(R)$ is the mass contained within the 
radius $R$, as originally shown by Blumenthal et al. (1986). 
This model, which assumes spherical symmetry and circular orbits, was improved by 
Gnedin et al. (2004). These authors showed that, when compared to numerical simulations, the
Blumethal et al. model overpredicts the increase of DM density in the central region and that the change of the 
assumed invariant from $M(R)R$ to $M(\bar{R})R$ (where $R$ and $\bar{R}$ are the current and
orbit-averaged particle positions) largely reduces the problem.

Gnedin et al. (2004) also estimate that, for $10^{-3}\lesssim(R/R_{\rm vir})\lesssim 1$,
\begin{equation} 
{{\bar{R}}\over{R_{\rm vir}}}
\simeq A \left({R\over{R_{\rm vir}}}\right)^w,
\label{r_bar_from_r}
\end{equation}
\noindent
with $A \sim 0.85 \pm 0.05$, $w \sim 0.8 \pm 0.2$. Gustaffson et al. (2006) confirmed these results, 
but showed that the values of $A$ and $w$ change from halo to halo, and that the spread is likely
larger than the errors quoted above. However, in the following we used
the central values ($A = 0.85$, $w = 0.8$) from Gnedin et al. (2004), as they
lie well within the distribution.

With such assumptions, the modified adiabatic invariant equation is
\begin{equation}
R_{\rm f}[M_{\rm DM,f}(\bar{R}_{\rm f})+M_{\rm b,f}(\bar{R}_{\rm f})] =
R_{\rm i}[M_{\rm DM,i}(\bar{R}_{\rm i})+M_{\rm b,i}(\bar{R}_{\rm i})]
\label{modified_adiabatic_invariant}
\end{equation}
\noindent
where $M_{\rm DM,f}(R)$, $M_{\rm b,f}(R)$, $M_{\rm DM,i}(R)$, and $M_{\rm b,i}(R)$ are
the masses of DM and baryons enclosed within a radius $R$, at the final
(subscript ${\rm f}$) and initial (subscript ${\rm i}$) times. Given the initial
profiles $M_{\rm DM,i}$ and $M_{\rm b,i}$ (the NFW profile described in Sec.
\ref{haloinimod}) and $M_{\rm b,f}$ (the baryonic density given by the stellar 
evolution models), Eq. (\ref{modified_adiabatic_invariant}) can be solved
iteratively for the radius $R_{\rm f}$ which encloses the DM mass $M_{\rm DM,i}(\bar{R}_{\rm i})$.

The numerical routine which solves the equation is mostly based on the
public code {\tt contra} by O. Gnedin\footnote{See {\tt http://www.astro.lsa.umich.edu/$\sim$ognedin/contra/}}, 
although several adaptations and changes were necessary.

Finally, it is important to  discuss our use of Eq. (\ref{r_bar_from_r})
down to $R \sim 10^{-7}R_{\rm vir}$, which is well below the limit ($\sim 10^{-3} R_{\rm vir}$) 
where it was tested by Gnedin et al. (2004). 
Although this is definitely an untested extrapolation, we think our choice is 
well motivated and quite conservative: the resulting central DM density is at 
least a factor of 10 lower than in the Blumenthal et al. (1986) model.
It is worth noting that our results are in agreement with what recently 
found by Freese et al. (2008b): they study the density profile of an AC contracted 
halo, adopting two different algorithms (based on Blumenthal's 
original prescription and on a modified method derived by Young (1980));
 they find the two to be consistent within a factor two at baryon densities of
${\cal O}(10^{-11})$g/cm$^3$, yielding DM densities of order 10$^{10}$GeV/cm$^3$ at 
radii of order 10 AU.
If we use the same initial conditions and apply the algorithm based on Gnedin's 
method to their baryonic profile,  we find a DM density consistent within a factor three 
with the ones they obtain. 

The specific energy deposition rate due to annihilations of DM particles is 

\begin{equation}
\frac{dL_{\rm DM}}{dV}=\frac{\rho^2}{m_\chi}\langle\sigma v\rangle;
\label{FSDMannih}
\end{equation}
\noindent
where $\rho$ is the local DM density, $m_\chi$ the neutralino mass, $\langle\sigma v\rangle$ the thermally--averaged 
annihilation rate. We adopt $\langle \sigma v \rangle = 3 \times 10^{-26}$~cm$^3$~s$^{-1}$ which 
best fits the current value of the DM relic density (Bertone et al. 2005). DM particles in the 
halo are strongly non--relativistic: therefore the p--wave term, which contributes to the annihilation rate in the 
early Universe, is negligible in astrophysical environments; this may lead to different values of 
$\langle\sigma v \rangle$. In general, however, the same value we adopt is taken as ``fiducial'' in DM indirect search 
studies (see e.g. Fornengo, Pieri \& Scopel 2004).
We also assume the neutralino mass to be $m_\chi = 100$~GeV; we will discuss the effects of the variation of these 
parameters in Sec. \ref{evolution}.
In general, only a fraction $f$ of the energy released in the annihilation is emitted in form of particles that can be 
thermalized by the gas; we take $f =2/3$, as for a typical neutralino annihilation $\approx 1/3$ of the energy goes 
into neutrinos \cite{Bertone:2004pz}, and the rest of hadronic and electromagnetic shower induced by the primaries 
fastly thermalizes inside the protostellar core, as estimated by Spolyar et al. (2008) for even lower baryonic densities\footnote{It is worth noting that the mean free path for electrons and photons with energy lower than the neutralino mass $m_\chi$, is much smaller than the radius of the star at any time during our analysis.}.

\subsection{Proto--star evolution with DM contraction}
\label{AdiaEvol}

We now turn to a detailed discussion of the DM effects on the evolution of a 100 $M_\odot$ proto--star, 
which we take to be our fiducial model deferring to Sec. \ref{evolution} the study of different masses. 
We anticipate that, qualitatively, the conclusions and the physical picture we draw in this Section do not 
depend strongly on the assumed stellar mass.

\begin{figure}
\centering
\includegraphics[angle=90,width=0.45\textwidth]{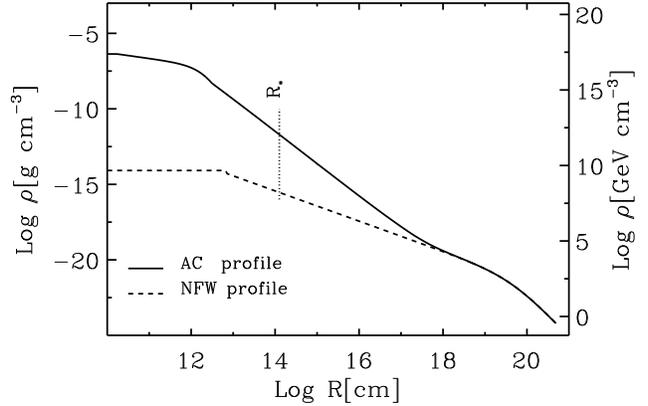}
\caption{Initial NFW DM density profile of adopted $M=10^6 M_\odot$ halo (dashed line) and adiabatically contracted 
DM profile at the time of our initial proto--stellar phase of for the fiducial 100 $M_{\odot}$ star. 
The vertical dashed line marks the radius of the star at the beginning of the computation.}
\label{DM_prof1}
\end{figure}

The initial NFW profile described in Sec. \ref{haloinimod} is adiabatically contracted, as described 
in the previous Section, in order to obtain a configuration in which the baryonic component would correspond to our 
initial proto--stellar phase, at the tip of the Hayashi track, presented in Sec. \ref{inistemod}. 
The initial NFW profile of the halo, 
and the AC contracted DM density distribution at the time we start our simulation are shown in Fig. \ref{DM_prof1}; 
the dramatic enhancement (approximately seven orders of magnitude within the central 10$^{12}$cm) of DM density is 
the cause of the pronounced effects of DM annihilation that will be discussed in the following.

The corresponding DM profile at this point is let evolve together with our stellar object, whose evolution follows a 
typical track in the HR diagram: the proto--star is totally convective and contracts on a Kelvin-Helmholtz time scale 
(approximately 10$^2$yr in this phase), descending the Hayashi track. For reference, the track of a 100 $M_\odot$ 
(together with other stellar masses) in the HR diagram is reported in Fig \ref{HR-38DMcap}.

The DM annihilation luminosity (AC luminosity) becomes the dominant component of the total luminosity on a very short 
timescale ($\approx$ 10 yr). Although the mass of DM contained inside the star at this point is only 
$\approx$10$^{31}$g, 10$^{-4}$ of its baryonic mass (and 10$^{-8}$ of the total mass of the halo), the luminosity 
arising from the annihilations of the DM concentrated within the central $\approx 2 \times 10^{11}$cm  is large 
enough to sustain the star, causing a {\it stalling phase}.
This kind of object has been named a {\it dark star} by Spolyar et al. (2008), who first found
that DM annihilation energy release can counteract the gravitational collapse at some point
during the pre--stellar phase.

Once the proto--stellar contraction has stalled, due to the energy released by DM annihilations, the contraction of 
the DM profile is inhibited as well. Part of the DM in the cusp is burned by annihilations causing a luminosity drop, 
followed by a small contraction of the baryons and DM which re-establishes a new equilibrium state.
In reality, this sequence of stable equilibrium states along the Hayashi track represents a continuous process which 
is eventually terminated when the DM density in the cusp has decreased below the threshold at which the energy input 
can no longer sustain the self-gravity pull, i.e when the annihilation timescale becomes longer than the local 
Kelvin-Helmholtz time. 
The duration of the stalling phase, $\tau_{\rm AC}$, 
is defined as the time needed to the AC luminosity to scale down to 50\% of the stellar luminosity, i.e. 
$L_{\rm DM}/L_{\ast} = 0.5$, and it is much longer than the typical Kelvin-Helmholtz timescale.
For our fiducial set of parameters (recall this is degenerate in the ratio $\langle \sigma v \rangle/m_\chi$), the 
100 $M_\odot$ dark star stalls for $\tau_{\rm AC} = 5.3 \times 10^3$yr. 

\begin{figure}
\centering
\includegraphics[angle=90,width=0.45\textwidth]{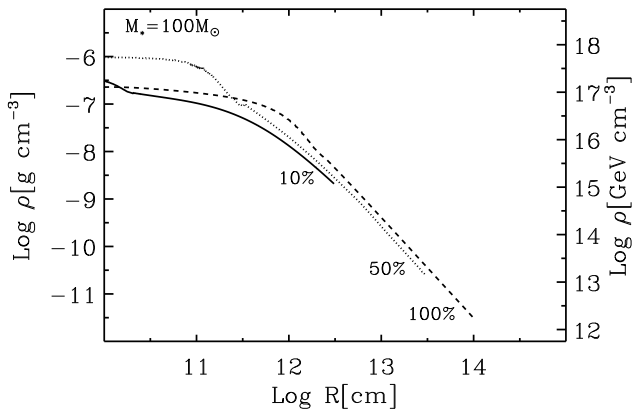}
\caption{DM density profile, truncated at the stellar radius, $R_\ast$, at different stages during the pre--MS evolution of 
the fiducial $100 M_\odot$ proto--star. Curves are labelled by the ratio $L_{\rm DM}/L_\ast$; conversion to 
corresponding time and temperatures can be read off Fig. \ref{DM_TH}.}
\label{DM_prof2}
\end{figure}

In Fig. \ref{DM_prof2} we show the DM density profile inside $R_\ast$ at different stages of the proto--stellar 
evolution, benchmarked by the fraction of stellar luminosity provided by AC luminosity, 
$L_{\rm DM}/L_\ast$. The contraction and subsequent flattening of the central cusp with respect to the 
initial conditions of the stalling phase, where $L_{\rm DM}/L_\ast = 1$, together with the progressive
shrinking and the loss of DM annihilating shells, is the reason of the efficiency decrease of the AC luminosity.
The central enhancement of the central cusp at the 50\% stage with respect to the 100\%, dark star phase, 
is not able to compensate the loss of the external shells of annihilating DM.

Fig. \ref{DM_TH} shows the time evolution of the effective temperature for different proto-stellar masses. 
During the stalling phase, the dark star is kept stable on the HR diagram by the DM energy production (which is temperature 
independent) and it remains cool $T_{\rm eff} \approx 5 \times 10^3$ K, as can be seen from the initial plateau visible in 
the bottom left corner of the Figure. 
When the DM cusp is exhausted, the dark star evolves along its pre--MS track with characteristic times longer than for a 
standard, non DM--supported, stellar model of the same mass, due to the additional support of AC luminosity. 
Fig. \ref{DM_TH} restates the results that the dark star exits the stalling phase in 
$\tau_{\rm AC} = 5.3 \times 10^3$~yr, and reaches the bottom of the Hayashi track in $\tau_{\rm Hay} = 4.4 \times 10^3$~yr.
By comparison a standard, non DM--supported, star of the same mass reaches the same point in $\tau_{\rm Hay}^0 \approx 60$~yr.
Table \ref{AdiaTable} shows these characteristic timescales for a grid of stellar mass models. 

This might have interesting implications for the radiative feedback effects of the dark star on its host and nearby
haloes, as we will discuss in our Conclusions. 

\section{Dark matter capture}
\label{capture}

If DM is made of weakly interacting massive particles, a non-vanishing elastic scattering cross section between 
DM and baryons arises. WIMPs scattering off the nuclei which constitute the star lose part of their energy and 
some of them remain bound by gravitational attraction. 
In this Section, we first introduce the formalism necessary to implement scattering/capture (SC) process in the code
and then we discuss their effects on the evolution of our $100 M_{\odot}$ fiducial proto--stellar model.

\subsection{Formalism and approximations}
\label{DMcapEq}

The capture rate, $C$, of DM particles by a star through scattering has been calculated by Gould (1987), and can be 
cast as follows:

\begin{equation}
C = 4 \pi \int_0^{R_\ast} dR \, R^2 \, \frac{dC(R)}{dV},
\label{Intcaprate}
\end{equation}
\noindent
where

\begin{eqnarray}
\frac{dC(R)}{dV} & = & \left(\frac{6}{\pi}\right)^{1/2}
\sigma_0 A_n^4 \frac{\rho_\ast}{M_n}\frac{\rho}{m_\chi}
\frac{v^2(R)}{\bar{v}^2} \frac{\bar{v}}{2\eta A^2} 
\label{Caprate}\\
& \times &
\left\{\left(A_+A_- -\frac12\right)
\left[\chi(-\eta,\eta) -\chi(A_-,A_+)\right]
\right.\nonumber\\
&+&\left.
\frac12A_+e^{-A_-^2} -\frac12A_- e^{-A_+^2}-\eta e^{-\eta^2}
\right\},\nonumber\\
A^2&=& \frac{3v^2(R)\mu}{2\bar{v}^2\mu_-^2},   \  \  A_\pm=A\pm\eta,  \  \   \eta=\sqrt \frac{3v_\ast^2}{2\bar{v}^2}, \nonumber\\
\chi(a,b)&=&\int_a^b dy\, e^{-y^2}=
\frac{\sqrt\pi}{2}[{\rm erf}(b)-{\rm erf}(a)],
\nonumber
\end{eqnarray}
\noindent
$\sigma_0$ is the DM-baryon elastic scattering cross section, $A_n$ ($M_n$) is the atomic number (mass) of stellar nuclei, 
$\rho$ is the ambient WIMP density, $\bar{v}$ is the WIMP velocity dispersion, $v_\ast$ is the velocity of the
star with respect to the observer, $v(R)$ is the escape velocity at a given radius $R$ inside the star, 
$\mu=m_\chi/M_n$, $\mu_-=(\mu-1)/2$, and the subscript $\ast$ refers to stellar quantities.
The factor $\eta$ is usually assumed to be $\sqrt{3/2}$, corresponding to the condition $v_\ast = \bar{v}$. 
WIMPs captured by the scattering process thermalize with the gas;
an upper limit to the thermalization time can be estimated as ($m_\chi/m_p$)($\lambda_\chi/v_\chi$), i.e. the 
WIMP-proton mass ratio (number of scatterings needed) times the WIMP mean free path divided by its dispersion 
velocity in the star (we take $v_\chi$ equal to the escape velocity at the stellar surface), thus obtaining:

\begin{equation}
\tau_{\rm th}=\frac{4\pi}{3\sqrt{2G}}\frac{m_\chi}{\sigma_0}\frac{R_\ast^{7/2}}{M_\ast^{3/2}}.
\label{tauth}
\end{equation}
\noindent
Assuming that the energy of the DM particles is in equilibrium with the gravitational field of the star, their number 
density follows a Maxwell-Boltzmann distribution \cite{gs87},   

\begin{equation}
n_{\chi}(R) =n^c_\chi\exp(-R^2/R_\chi^2), \ \  n_\chi^c=\frac{C \tau_\chi}{\pi^{3/2}R^3_\chi};
\label{DMprofile} 
\end{equation}
\noindent
where $n_\chi^c$ is the highest DM density achievable inside the star once equilibrium between capture and 
annihilation is reached after a time

\begin{equation}
\tau_{\chi}=\left(\frac{\pi^{3/2}R^3_\chi}{C\langle\sigma v\rangle}\right)^{1/2}.\label{Eqtimescale} 
\end{equation}
\noindent
The radius within which captured DM is concentrated, once it has thermalized with the star, reads

\begin{equation}
R_{\chi}=c\left(\frac{3kT_c}{2\pi G\rho_c m_\chi}\right)^{1/2}, \label{radDMeq} 
\end{equation}
\noindent
where $T_c$ and $\rho_c$ are the stellar core temperature and density, respectively. 

The energy released due to annihilations {\it inside} the star can be self--consistently computed once the 
profile $n_\chi(R)$ is known. The expression for such quantity given in Eq. (\ref{DMprofile}) assumes that 
particles are thermalized and equilibrium between annihilation and capture processes has been reached. 
To take into account the transient phase before WIMPs settle to such a state, we write the annihilation 
luminosity as,  

\begin{equation}
L_{\rm DM}(t)=g(t)~4\pi f \int_0^{R_*}dR\ R^2 n^2_\chi(R) \langle\sigma v\rangle  m_\chi,
\label{integLumprof}
\end{equation}
\noindent
with $g(t)=\tanh^2(t/\tau_{\rm dyn})$, where $\tau_{\rm dyn}$=max($\tau_\chi$,$\tau_{\rm th}$); this is a formal 
solution when $\tau_\chi >\tau_{\rm th}$, and otherwise still represents a good approximation to deal with a 
transient, reducing to the exact solution, $L_{\rm DM}=C m_\chi f$, when $t > \tau_{\rm th}$. Finally, $f$ is the 
fraction of released energy absorbed within the star, which we take to be 2/3 for a typical neutralino 
annihilation, as discussed in Sec. \ref{adiabatic}.

The process we have described presents some peculiarities which are worth discussing. Altough the physical energy 
source is the annihilation process, its rate is controlled by scattering processes which governs the capture rate. 
Its dependence on the background WIMP density is only linear (rather than quadratic, as in the case of annihilation 
reactions), and it depends on $\sigma_0$ rather than on $\langle\sigma v\rangle$. 
As it can be appreciated from Eqs. (\ref{DMprofile}) and (\ref{Eqtimescale}), $\langle\sigma v\rangle$ and 
$m_\chi$ affect the $\tau_\chi$ and $n^c_\chi$, but within the region of the parameter space which is relevant
to this problem they do not affect the final DM luminosity.

Eq. (\ref{Caprate}) should be integrated for each single atomic species in the star. However, if one relies on the 
current experimental upper limits for DM direct detection for a 100 GeV mass neutralino, namely 
$\sigma_0^{\rm d}$=10$^{-38}$cm$^2$ (Desai et al. 2004\footnote{SuperKamiokande Collaboration} and Angle et al. 2008\footnote{XENON10 Collaboration}), 
and $\sigma_0^{\rm i} =4 \times 10^{-44}$cm$^2$ (Ahmed et al. 2008\footnote{CDMS Collaboration}), 
the capture rate is negligible for any species but hydrogen. 
In stars of primordial composition, such as Pop III ones, even the dependence on the coherence factor $A_n^4$ 
does not introduce any signicant contribution of elements other than hydrogen.
Our choice for the cross section values is in agreement with other works in the literature, i.~e.~ 
\cite{Moskalenko:2007ak,Bertone:2007ae,Fairbairn:2007bn}.

Throughout the following we adopt $\bar v=10$~km~s$^{-1}$, which represents the virial velocity of our reference 
mini--halo with mass $10^6 M_\odot$ at redshift $z \approx 20$. 

By integrating Eq. (\ref{Caprate}) with a flat stellar density profile
, one obtains a simplified expression for the capture rate,

\begin{equation}
C \propto   \sigma_0 M_\ast v_{\rm esc}^2 \frac{\rho}{m_\chi}= C_0 \sigma_0 \frac{M^2_\ast}{R_\ast}
\frac{\rho}{m_\chi}.
\label{propcap}
\end{equation}
\noindent
which, within the precision of experimental data, corresponds to a numerical estimate of,

\begin{equation}
C = 9.2 \times 10^{47}{\rm s}^{-1}\frac{M^2_\ast}{R_\ast} \frac{\rho_{11}\sigma_{38}}{m_{100}},
\label{paramCapt}
\end{equation}
\noindent
having defined:

\begin{equation}
\rho_{11}=\frac{\rho}{10^{11}\frac{\rm GeV}{{\rm cm}^{3}}};~ \sigma_{38}=\frac{\sigma_0}{10^{-38}{\rm cm}^2};~m_{100}=\frac{m_\chi}{100{\rm GeV}},
\end{equation}
and expressing $M_\ast$ in solar masses and $R_\ast$ in cm.

It follows that, at equilibrium,
\begin{equation}
L_{\rm DM} = 1.4 \times 10^{47}\frac{\rm erg}{\rm s}\frac{M^2_\ast}{R_\ast} \frac{\rho_{11}\sigma_{38}}{m_{100}}.
\label{paramLum}
\end{equation}
\noindent
In the above expression we have taken $f=2/3$ (see discussed in Sec. \ref{adiabatic}). Eq. (\ref{paramLum}) predicts 
that for a given mass $M{_\ast}$, $L_{\rm DM}$ will grow during stellar contraction, potentially reaching a level able to 
halt the collapse. In the code, we have implemented a term of DM luminosity due to annihilations of captured WIMPs (SC luminosity)
using Eq. (\ref{paramLum}) multiplied by the transient factor $g(t)$ described in Eq. (\ref{integLumprof}). 

We emphasize that the process of DM capture is sensitive to the background DM density outside the star but 
not to the DM already accreted through adiabatic contraction. Thus, the two processes are mutually 
independent. Moreover, scattering/capture processes can continue for long times: in a $10^6 M_{\odot}$ 
halo, a DM luminosity of $10^{41}$~erg~s$^{-1}$ can be sustained for approximately 10$^{12}$~yr.

\subsection{Proto-star evolution with DM capture}
\label{DMcapEvol}

We follow the evolution of our reference $100 \, M_\odot$ proto--star, after the stalling phase due to the 
AC luminosity described in Sec. \ref{AdiaEvol}.
The fiducial DM parameters are $\sigma_0 =10^{-38}$cm$^2$ and $\rho =10^{11}$GeV/cm$^3$, which is the value set by 
adiabatic contraction in the vicinity of the star ($R \lesssim 10^{15}$~cm), as shown in Fig. \ref{DM_prof1}.  
It is worth noticing that this value of $\rho$ closely matches the one predicted by 3D simulations of first star formation 
\cite{TurkFS3}. We will discuss the dependence of our model results on these parameter in Sec. \ref{evolution}. 

At the time of stallation, the radius of the dark star is $R_{\ast} \approx 10^{14}$cm. Once the DM density cusp 
generated through AC is exhausted, the energy released by DM annihilations is no longer sufficient to stop the 
gravitational collapse; also, the SC luminosity developed at this point is, at the 
equilibrium\footnote{We note that $\tau_{th}\approx10^{10}$yr at this stage, thus making the actual SC luminosity 
even smaller that its equilibrium value.}, approximately $10^{37}$ erg/s, as it can be calculated with Eq. (\ref{paramLum}).
The dark star continues its evolution along the Hayashi line and in the pre--MS phase. 

\begin{figure}
\centering
\includegraphics[angle=90,width=0.45\textwidth]{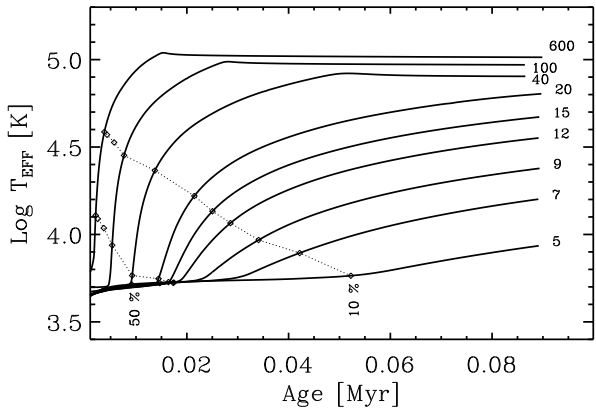}
\caption{Evolution of the effective temperature for a selected set of stellar models when only the effects of AC DM annihilations
is considered. The dotted lines indicate the evolutionary stages of the stars when the AC DM luminosity has decreased to
50\% and 10\% of the total stellar luminosity.}
\label{DM_TH}
\end{figure}

\begin{figure}
\includegraphics[angle=90,width=0.45\textwidth]{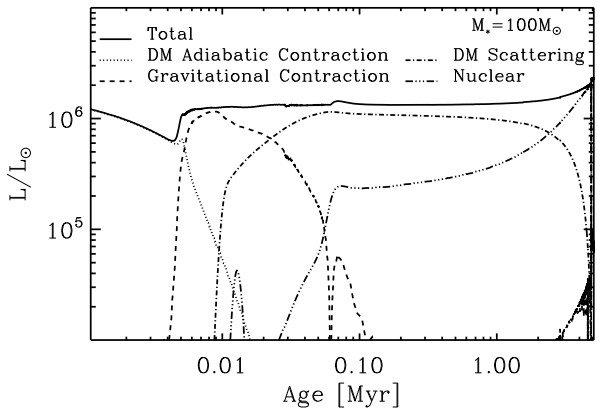}
\caption{Total luminosity of our reference $100 M_{\odot}$ star as a function of its age. The different curves show
the contribution from AC DM annihilations (dotted line), gravitational contraction (short dashed), SC DM annihilations
(dot-dashed), nuclear reactions (dot-dot-dot-dashed). Where relevant, the processes have been computed for our 
fiducial values of $\sigma_0$ and $\rho$ (see text).}
\label{evolEner1}
\end{figure}

While the star shrinks and evolves leftward on its track in the HR diagram, the capture rate grows as can be seen in 
Fig. \ref{evolEner1}, where we show the contributions to the total stellar luminosity of the different processes as a 
function of the stellar age. 
As shown in the Figure, despite the fact that SC DM annihilations are dominating the overall luminosity,
at times $>$2$\times 10^3$ yr, nuclear reactions are active producing a luminosity 
of $L_{\rm nucl} \approx 2 \times 10^{38}$erg/s ($10^5 L_\odot$), eventually 
leading the star to exhaust the hydrogen into its core and continue its evolution, although with longer timescales.
We have followed the evolution of our fiducial stellar model until complete consumption of helium in the stellar core. 
The hydrogen burning lifetime is $\tau_{\rm H}= 4.9$~Myr, to be compared with 
the $\tau_{\rm H}^0 = 2.6$~Myr predicted for a Pop III star of the 
same mass in the absence of DM effects (see Table \ref{SCTab}).
The helium burning lifetime remains essentially unchanged.

Thus, the evolution of the star is slower than what expected for a star of the same mass in the
absence of DM capture.

As we will show in the next Section, lower mass stars are more sensitive to DM effects and for this same set of parameters 
they actually stop before getting to the ZAMS.

\section{Parameter variation}
\label{evolution}

In this Section we will discuss the dependence of model results on the assumed parameters. In particular, we
will explore (i) a grid of stellar masses and (ii) different DM parameters. We will consider these in turn.

We find that all stars stall very early in their evolution, when they all
stand on the Hayashi track  (namely, they are entirely convective). 
The evolution of the effective temperature during the AC phase for a few selected stellar models is shown 
in Fig. \ref{DM_TH}, where the dotted lines mark the evolutionary stages where the AC luminosity has 
decreased to 50\% and 10\% of the total stellar luminosity. As can be inferred from the Figure, larger masses 
burn their DM content more rapidly than smaller ones.
In Table \ref{AdiaTable}, we report the characteristic timescale of the AC phase and the time taken by each star
to reach the bottom of the Hayashi track  with ($\tau_{\rm Hay}$) and without ($\tau_{\rm Hay}^0$) the contribution 
of AC luminosity for the whole range of stellar masses under investigation.
The duration of the stalling phase induced by the AC luminosity varies with stellar mass M$_\ast$, ranging from  
$2.1 \times 10^3$yr for a $600 M_\odot$ star to $1.8 \times 10^4$yr for a $9 M_\odot$ star.  

We have not been able to explore a wide range of values for $\langle\sigma v\rangle$/$m_\chi$ for problems of 
numerical convergence; however, we have performed a run for our fiducial value of the annihilation rate and a 
neutralino mass $m_\chi$ = 200 GeV. Results are reported in the Appendix \ref{datatab}, and show that for 
higher values of the neutralino mass (or smaller values of the annihilation rate) the delaying effect of the 
AC luminosity is reduced: the smaller amount of energy per unit time makes the contraction of the star faster.

\begin{figure}
\centering
\includegraphics[angle=90,width=0.45\textwidth]{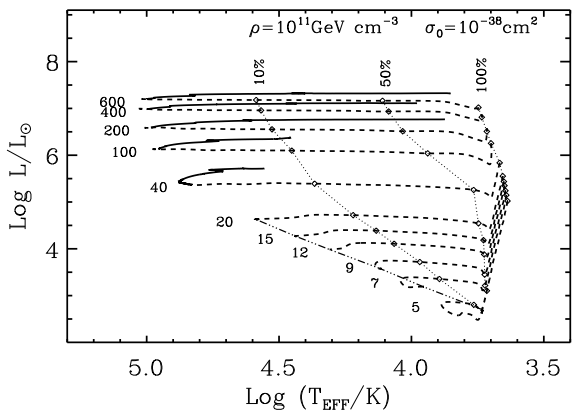}
\includegraphics[angle=90,width=0.45\textwidth]{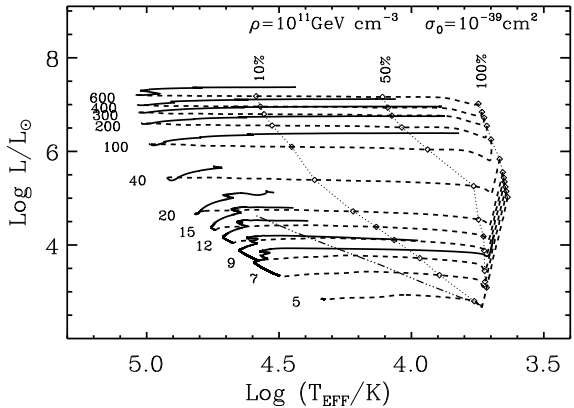}
\caption{The HR diagram for a grid of stellar masses. For each stellar model, the dashed line represents the pre--MS phase and 
the solid line represent the MS. Dotted diagonal lines mark the evolutionary stages when $L_{\rm AC}/L_\ast = 1$, 0.5 and 0.1. 
The dot--dashed line illustrate the locus of the ``freezing'' points, when the evolution is halted by SC DM annhilation luminosity
before the ZAMS. In the {\it upper} panel, the results have been obtained using our fiducial DM parameters, namely
$\sigma_0 = 10^{-38}$cm$^2$ and $\rho = 10^{11}$GeV cm$^{-3}$. 
The small loops in the 5 and 7 M$_\odot$ models are due to the effect of WIMPs thermalization, which results in an 
effective delay of the effects of SC DM annihilation luminosity.
In the {\it lower} panel, the stellar models have been run 
assuming the same DM density but $\sigma_0 = 10^{-39}$cm$^2$.}
\label{HR-38DMcap}
\end{figure}

Fig. \ref{HR-38DMcap} shows the evolution of different stellar models in the HR diagram. The results presented in the
upper panel have been obtained using our fiducial DM parameters, namely $\sigma_0 = 10^{-38}$cm$^2$ and 
$\rho=10^{11}$GeV cm$^{-3}$. The dotted lines mark the position of the star when $L_{\rm DM}/L_\ast$ = 1 (i.e. at the
beginning of the stalling phase), 0.5 and 0.1. 
As already discussed, for all the considered stellar models the stalling phase takes place along the Hayashi 
track, and the timing of the other benchmarked points can be inferred by comparison with Fig. \ref{DM_TH}.

\begin{table*}
\centering
\caption{Characteristic times of the star relative to the phase induced by the AC luminosity.
The adiabatic time $\tau_{\rm AC}$  has been defined as the time needed to the AC DM annihilation luminosity, $L_{\rm DM}$ 
to scale down to 50\% of the total stellar luminosity $L_\ast$. 
We also report the stellar radii $R_{\rm AC}$. The last two columns show the time required to reach the bottom of the 
Hayashi track from its tip with ($\tau_{\rm Hay}$) and without ($\tau^{\rm 0}_{\rm Hay}$) AC DM annihilation.}
\begin{tabular}{@{}ccccc@{}}
\hline
        M(M$_\odot$)  &   $\tau_{\rm AC}$($10^3$yr)&   $R_{\rm AC}$(cm)  & $\tau_{\rm Hay}$(yr) &$\tau^{\rm 0}_{\rm Hay}$(yr)\\
\hline

        5  & 9.7 &    4.0$\times$10$^{13}$&   3.2$\times$10$^{4}$ & 1.8$\times$10$^4$ \\
         7   & 15 &     4.6$\times$10$^{13}$&    2.1$\times$10$^{4}$ & 6.9$\times$10$^3$\\
         9   & 18   &    5.0$\times$10$^{13}$&   1.8$\times$10$^{4}$ &3.5$\times$10$^3$\\  
        12  &  18  & 5.6$\times$10$^{13}$&    1.5$\times$10$^{4}$ &   1.6$\times$10$^3$\\
        15   & 16   &   6.1$\times$10$^{13}$&  1.4$\times$10$^{4}$ &  1.0$\times$10$^3$\\
        20   & 14   &      6.8$\times$10$^{13}$&  1.2$\times$10$^{4}$ &   5.6$\times$10$^2$\\
        40   & 9.3   &      9.0$\times$10$^{13}$&   8.2$\times$10$^{3}$ &   2.0$\times$10$^2$ \\
       100   & 5.3 &       1.2$\times$10$^{14}$&    4.4$\times$10$^{3}$ & 62 \\
       200    & 3.7 &     1.6$\times$10$^{14}$&  2.5$\times$10$^2$ &   26 \\
       400  &  2.5  &     2.1$\times$10$^{14}$&  95 &  13 \\
       600   & 2.1  &   2.4$\times$10$^{14}$&   45 &  3.9\\
\hline
\end{tabular}
\label{AdiaTable}
\end{table*}

Following the stalling phase, these dark stars continue their contraction. 
The SC mechanism becomes more relevant as the density increases, and the SC rate grows.
For our fiducial set of DM parameters, stars with $M_{\ast} \lesssim 30 M_\odot$ develop a SC luminosity 
greater than the gravitational one before reaching the ZAMS, and therefore do not evolve 
further on the HR diagram.
This is clearly shown in the upper panel of Fig. \ref{HR-38DMcap}, where the dashed tracks, which
indicate the pre--MS phase, do not join the main sequence (solid lines) for all stellar models with 
$M_\ast \leq 20 M_\odot$. The lower panel of the Figure shows the HR diagram of the
same set of stellar models but assuming different DM parameters, namely $\sigma_0$=10$^{-39}$cm$^2$
and $\rho=10^{11}$GeV cm$^{-3}$. As expected, the evolution is almost unaffected by the 
variation of DM parameters during the AC phase whereas the SC effects are significantly reduced: 
all the stars with $M_\ast > 5 M_\odot$, reach the main sequence and even the evolution of the 
$5 M_{\odot}$ model is halted at a later time with respect to the previous case. 

In Table \ref{SCTab}, we summarize the characteristic timescales which characterize the evolution of
the our stellar models under the effect of SC assuming different DM cross sections, as compared 
with the standard case where DM effects are neglected: once again it is evident the bigger impact of 
the SC mechanism on smaller masses and the life--prolonging effect of DM on all the masses. 

\begin{table*}
\centering
\caption{Times needed to nuclear luminosity $L_{\rm nucl}$ to be 95\% of the total L$_\ast$, $\tau_{\rm Z}$; 
$\tau_{\rm H}$ is the time at which the hydrogen is totally exhausted in the core; underscript $0$ refers to 
the case of complete absence of any dark matter annihilation mechanism. $\sigma_{38}$($\sigma_{39}$)=10$^{-38}$(10$^{-39}$) cm$^2$, $\rho$=10$^{11}$GeV/cm$^3$.}

\begin{tabular}{@{}ccccccc@{}}
\hline
        M(M$_\odot$)&$\tau^0_{Z}$($10^3$yr)&$\tau_Z$($10^3$yr)($\sigma_{39}$)&$\tau_Z$($10^3$yr)($\sigma_{38}$)&$\tau^0_{H}$(Myr)&$\tau_H$(Myr)($\sigma_{39}$)&$\tau_H$(Myr)($\sigma_{38}$)\\
\hline
         7  &  4.9  $\times$10$^2$&   5.8 $\times$10$^4$ & Stalling & 29& 60& Stalling\\
         9  &  4.0  $\times$10$^2$&  3.1$\times$10$^4$ & Stalling & 20& 33& Stalling\\
        12 &   2.0$\times$10$^2$&  1.6$\times$10$^4$ & Stalling & 14 & 18 &   Stalling\\
        15 & 1.5$\times$10$^2$ &1.1$\times$10$^4$ &Stalling &11&13 & Stalling\\
        20  &  1.0 $\times$10$^2$&  6.4$\times$10$^3$& Stalling & 7.7 &  8.6&Stalling\\
        40  &  41&   2.4$\times$10$^3$ &2.1$\times$10$^4$& 4.1& 4.4& 21\\
       100 &   22&   1.1$\times$10$^3$&  4.6$\times$10$^3$& 2.6& 3.0 &4.9\\
       200  &   17&  9.6$\times$10$^2$& 3.6$\times$10$^3$& 2.2 & 2.4   & 3.9\\
       400 & 14& 1.1$\times$10$^3$& 3.4$\times$10$^3$&2.0 & 2.0& 3.7\\
       600  &  13 &  9.7$\times$10$^2$& 4.0$\times$10$^3$&  2.1&2.0 & 4.1\\
  \hline
  \end{tabular}
  \label{SCTab}
 \end{table*}

\section{Stellar mass constraints}
\label{DMconstrSF}

\begin{figure}
\includegraphics[angle=90,width=0.45\textwidth]{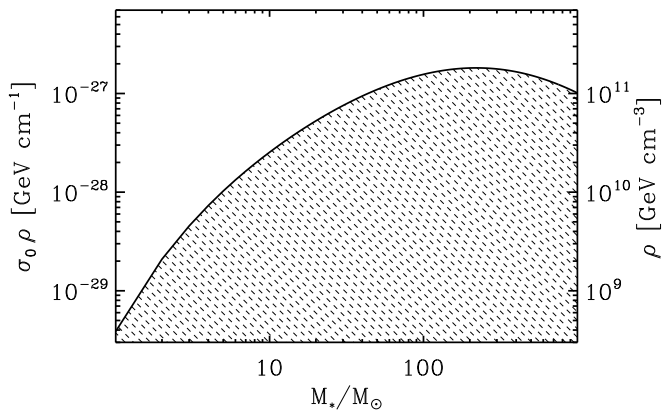}
\caption{Dark matter constraints on stellar mass. The shaded area represents the region of the parameter space where stars can 
reach the ZAMS and evolve. The vertical axis on the left represents the quantity $\sigma_0 \rho$ and on the right the corresponding 
values of $\rho$ if $\sigma_0 = 10^{-38}$cm$^2$. In the region above the curve, stars are prevented from reaching ZAMS and freeze on 
the HR diagram, as it is shown in Fig. \ref{HR-38DMcap}.}
\label{sigrhoDMconstr}
\end{figure}

From the above picture, it is clear that in DM-rich environments, such as the haloes where the first episodes of star 
formation are expected to take place, SC luminosity may play a dramatic role during the early evolution of a proto--star.
On the basis of purely quantitative arguments, this has been suggested by Iocco (2008), and  
Freese et al. (2008) derived a constraint on the mass of stars that can form under these conditions. 
Our analysis reaches conclusions that are different from those of the latter study as we will comment later in this Section. 
Now we want to answer the question of which stars will be most affected by this mechanism, and in which environment.
Armed with the formalism developed in Sec. \ref{capture}, one may simply impose the condition 
$L_{\rm DM}^{\rm ZAMS} \leq L_{\rm nucl}^{\rm ZAMS}$, namely that the luminosity due to DM annihilations inside the star be 
less than the nuclear luminosity predicted at ZAMS for Pop III stars in the absence of dark matter. 
Fig. \ref{sigrhoDMconstr} summarizes the results of this disequation, obtained 
by applying Eq. (\ref{paramLum}) to a grid of stellar models at the ZAMS.

In the region above the curve, DM luminosity exceeds the nuclear one and proto--stellar evolution is inhibited before 
the objects reach the ZAMS, as we have discussed in the previous Section. 
Stars in this regime will ``freeze'' on the HR diagram as long as the properties
of the DM distribution around them remain the same.
Below the curve (shaded area) stars are instead able to reach the ZAMS and 
thereafter evolve along the main sequence. 

Note, however, that the distinction between ``frozen'' and ``evolving'' stars has to be taken with care. In fact,
as we have shown in Fig. \ref{evolEner1} for our reference $100 \, M_\odot$ stellar model, the ignition of nuclear 
reactions can occur at very low rates, producing a very low nuclear luminosity, but still allowing the star to    
evolve, eventually exhausting its nuclear fuel, although on much longer timescales ($\tau_{\rm H} = 4.9$ Myr 
against the $\tau_{\rm H}^0 = 2.6$ Myr timescale predicted for a 100 $M_{\odot}$ Pop III star in the absence of 
DM effects, see Table \ref{SCTab}).

Fig. \ref{sigrhoDMconstr} has to be considered as a quantitative indication: it shows that both low and high mass stars 
are affected by DM capture. The range of stellar masses that can reach the ZAMS, for the fiducial
DM parameter combination ($\sigma_0 \rho = 10^{-27}$ GeV/cm), is $40 M_{\odot} \lesssim  M_\ast \lesssim 
1000 M_{\odot}$. 

This is due to a change in the index $k$ of the relation $L_{\ast} \propto  M_\ast^k$, following the transition 
of the star to a completely adiabatic system when $M_\ast \gtrsim 200 M_\odot$.
In their analysis, Freese et al. (2008) derive only an upper limit to the stellar mass in the presence
of DM SC because they impose the above condition making use of the Eddington luminosity, which has a 
linear dependence on the stellar mass, instead of the actual nuclear luminosity at ZAMS. 
Therefore their analysis somehow under--predicts the effects of DM on low and intermediate mass stars.
In fact, our numerical results show that most significant effects are obtained for objects with 
masses $M_{\ast}\lesssim$200M$_\odot$, as shown in Fig. \ref{sigrhoDMconstr}.
This happens because the SC mechanism becomes efficient only when the stars are in a relatively
advanced phase of their pre-MS evolution. 

It is also extremely interesting to notice that the mass--luminosity relation in low-mass stars is not 
particularly sensitive to metallicity. This implies that the results we have presented can be also
applied to metal--enriched stellar populations, as long as they continue to form in environments with
characteristics similar to those we have considered (see Sec. \ref{inimodel}).
Our results are in agreement with those of Salati \& Silk (1989), Fairbairn et al. (2008)
and Scott et al. (2007), who studied the main sequence evolution of WIMP--burning stars and observed
that ZAMS stars, if ``fed'' with captured DM annihilation energy move towards the red region of the HR 
diagram, at increasing DM densities.
In particular, the latter analysis focused on low--mass stars ($M_{\ast} \leq 4M_\odot$) 
and found they eventually reach the Hayashi line for different DM 
density values at different masses ($\rho = 10^{10}$GeV/cm$^3$ for 1 $M_\odot$),
using the same current upper limit on $\sigma_0$ we adopt, and a different value
for $\bar v$, of the order of the relative velocity between WIMPs and Sun, $\bar v \sim{\cal O}$(10$^2$Km/s).

\section{Conclusions}
\label{Concl}

We have studied the effects of WIMP dark matter annihilation on the first stars in the Universe.
As initial condition of our model, we consider a dark matter halo with mass $10^6 M_{\odot}$ 
at $z = 20$, i.e. a typical mini--halo where the first star formation episodes are expected to occur.
We have treated separately the mechanism of adiabatic contraction (AC) and scattering/capture 
(SC) and highlighted their effects on the pre--Main Sequence (pre--MS) phase of stellar objects 
with masses $5 M_{\odot} < M_{\ast} < 600 M_{\odot}$ formed from the collapse of metal--free 
gas clouds. We find that:

\begin{itemize}

\item Early in the proto-stellar evolution, the luminosity produced by DM annhilations during the 
AC regime induces a transient stalling phase: all the proto--stars become {\it dark} stars for 
characteristic times ranging from $2.1 \times 10^3$yr for a $600 M_\odot$ star to 
$1.8 \times 10^4$yr for a $9 M_\odot$ star.

\item The stalling phase occurs when the stars are on the Hayashi track. 
The AC luminosity does moderate the effective temperature of the star, 
by enabling an equilibrium state at the early evolutionary stages, characterized by 
larger radii, when the effective temperature is $\approx$5$\times$10$^3$ K.

\end{itemize}

Later in the evolution, the capture of WIMPs by means of scattering on the baryonic matter of the 
star becomes high enough that WIMPs accumulated and annihilating inside it produce enough energy 
to keep the object at equilibrium. The details of this depend on environmental conditions of DM, 
elastic scattering cross section between WIMPs and baryons, and mass of the star.

For our fiducial set of parameters ($\bar v$=10 km/s, $\rho$=10$^{11}$GeV/cm$^3$, $\sigma_0$=10$^{-38}$cm$^2$),
we find that:

\begin{itemize}

\item  Small and intermediate mass stars ($ M_{\ast} < 40 M_\odot$) are most affected by SC luminosity and their
evolution is halted on the HR diagram before reaching the ZAMS. 

\item {\it Dark} stars can be supported by SC luminosity as long as the environmental conditions remain unaltered. This
is because, unlike the AC mechanism, SC luminosity depends on the flux of 
DM particles streaming through the star from outside, thus drawing from a virtually unexhaustable reservoir: 
{\it dark} stars remain ``frozen'' on the HR diagram.

\item Stars with  masses $\geq 40 M_\odot$ manage to ignite nuclear reactions, and go through the 
Main Sequence supported by an additional energy source: dark matter ``burning'' prolonges their lifetimes from a
factor 2 for a $600 M_{\odot}$ to a factor 5 for a $40 M_{\odot}$ star.

\end{itemize}

These conclusions depend on the assumed dark matter parameters and on the specific environment where the 
first stars are expected to form.  
However, they do not strongly depend on the assumed primordial chemical composition of the stars. 
Thus, they can be applied to more evolved stellar populations as long as the characteristics of the environment where 
these stars form are similar to the ones we have considered.

The existence of {\it dark} stars can have many interesting consequences for a number of issues. 
In fact, once they are frozen on the HR diagram, they have effective temperatures in the range $10^4 - 
10^5$ K and provide a continuous source of UV photons. This could have interesting consequences for 
the radiative feedback on the parent and neighbouring haloes as well as on the reionization history. 
The duration of this {\it dark} stellar phase is difficult to estimate, as it depends on the persistence 
of high dark matter densities around the stars. Therefore, their fate is strictly
related to the evolution of their parent dark matter haloes and their merger histories.    

Of course, the present analysis represents only a first step of more refined future studies; however,
it opens a relatively novel window on high redshift star formation. Progresses on
the issues discussed here might lead to considerable understanding of the signatures that the yet
mysterious dark matter particles have unmistakably left on the stars that formed in the baby Universe.

\section*{Acknowledgements}
F.~I. acknowledges support from the grant COFIN--MIUR Pacini2006. A.~B. and P.~M. acknowledge financial contribution from contract ASI I/016/07/0.

We thank  G.~Bertelli, L.~Girardi, and M.~Mapelli for stimulating discussions.
F.~I. thanks M.~Fairbairn and P.~Scott for useful conversations.

\appendix
\section{Stellar evolution}

\subsection{The code}
\label{StelCoDet}

The modifications introduced in the Padova stellar  code
to follow the evolution of zero-metallicity stars
are fully described in Marigo et al. (2001, 2003)
to which the reader is referred for details.

In the following we summarize the main input physics
and further implementation required to follow the 
pre-main sequence phase.

Radiative opacities are from the OPAL group, Rogers \& Iglesias (1992) and
Iglesias \& Rogers (1993), for temperatures T $\geq$10$^4$K,
and from Alexander \& Ferguson (1994) for T $<$10$^4$K.
Conductive opacities of electron-degenerate matter are from 
Itoh et al.(1983).

The equation of state for temperatures higher than $10^7$~K is that of a fully-ionized
gas. At high densities, Coulomb interactions are
introduced adopting the prescription by Straniero (1988)

Nuclear reaction rates are from the compilation of Caughlan \& Fowler (1988) while,
energy losses by pair, plasma, and bremsstrahlung neutrinos, are from 
Haft, Raffelt \& Weiss (1995).

The energy transport in the outer convection zone is described according 
to the mixing-length theory of B\"ohm-Vitense (1958) with a 
mixing length parameter, $\alpha=1.68$, obtained from the calibration
of the solar model by Girardi et al. (2000).

The extension of convective boundaries is estimated with the standard 
Schwarzschild criterion. Adopting more sophisticated schemes such as
semiconvection and/or convective overshoot (e.g. Bressan et al. 1981
and Alongi et al 1991) would not change the results.

Effects of stellar rotation and/or magnetic fields have not been 
considered in this exploratory work.

We finally remark that for zero-metallicity stars 
the abundance equations need to be solved 
simultaneously, for both the H and He burning reactions
without any assumption for nuclear equilibria.
This is performed with a semi-implicit extrapolation
scheme (Bader \& Deuflhard 1983).

\subsection{Evolutionary Tracks}
\label{datatab}

We have computed evolutionary tracks  
for initial masses in the range 5 M$_\odot$ to 600 M$_\odot$, starting from the
pre-main sequence phase and covering the central Hydrogen and Helium burning phases.
The adopted initial composition consists of a mixture of just hydrogen and helium,
resembling the lack of metals in the early Universe \cite{Iocco:2007km}, with 
mass fractions of $X=0.755$ and $Y=0.245$, respectively.

The evolution is performed at constant mass, i.e. neglecting both mass accretion during 
the pre-main sequence phase and mass loss by stellar winds in the later stages.

Five different sets of tracks have been considered, depending on the assumptions concerning the DM 
parameters. The results are summarized in Tables A1 to A5. 
For selected stages during the  evolution we report: age, surface luminosity $L$, effective temperature $T_{eff}$, 
radius $R_*$, central values of density $\rho_c$, temperature $T_c$ and hydrogen fraction  $X_c$, and the 
fractional luminosity provided by nuclear reactions (L$_{Nuc}$/L$_*$).

Table A1 refers to stellar evolution without
DM effects (the standard case). 
In Table A2 we consider only the effects of annihilation of DM
in the adiabatic contraction phase. Tracks in tables 
A3 and A4 also take into account
the additional role of DM scattering/capture processes 
for two different values of the parameter $\sigma_0$, $\sigma_0=10^{-39}$cm$^2$ 
and $\sigma_0=10^{-38}$cm$^2$.
Finally in Table A5 we have analysed the influence of a different 
neutralino mass (200 GeV instead of 100 GeV) on the adiabatic contraction phase. 

The different stages selected in the Tables have the following meaning.

The {\sl Starting Point} marks the beginning of the evolutionary sequences.
In the standard case this is the first point where the model is fully and
consistently supported by the gravitational energy release, while in the AC
cases it is the model where the total luminosity of the star is equalized by
the AC luminosity of its current baryonic configuration. At these luminosities
the contraction timescale are of the order of a few years, so that
some differences
in the starting luminosity do not affect the subsequent evolutionary timescales.

The {\sl Hayashi Minimum Luminosity Model} corresponds to the stage of minimum luminosity
during the descent along the Hayashi locus. Soon after this point
the proto-star moves towards the main sequence at nearly constant luminosity.
This is a convenient point for comparing the contraction timescales 
of the different sets and their sensitivity to DM adiabatic contraction.

The points labelled {L$_{AC}$/L$_{*}$=50\% and 10\%} indicate the stages
when the AC luminosity contributes to the 50\% and 10\%, respectively, of the total
stellar luminosity. The complementary energy source comes from gravitational 
contraction. When L$_{AC}$/L$_{*}$=10\%, the adiabatic contraction 
phase is essentially over.

The stage L$_{Nuc.}$/L$_{*}$= 95\% indicate the point where the nuclear
energy source provides 95\% of the stellar luminosity.
In absence of DM scattering/capture this point indicates the 
beginning of the major Hydrogen nuclear burning phase. 
On the contrary when DM scattering/capture becomes efficient,
this point may be reached at an advanced stage of nuclear burning,
or even not reached at all, for stars that suffer SC stalling.
The contribution of the SC energy source makes the nuclear
burning to occur at a slower rate, thus prolonging
the lifetime of the star by a significant amount.
In some circumstances the star is practically totally sustained by the SC
luminosity with a negligible nuclear burning.
In all these cases we provide the central hydrogen fraction and the 
fractional nuclear luminosity. These quantities may help the reader 
to evaluate the importance of the effect and, for the stalled
stars, to roughly estimate the duration of the phase.

Finally, the last stage in the tables corresponds either to  central Hydrogen 
exhaustion, or to the stalled model at an age of 5 Myr. 
We remark that, in the latter case, the stellar tracks 
have been evolved for a much longer time (larger than
the standard Hydrogen burning lifetime) to properly check the stalling condition.
In Tables A3 and A4 we do not report the first stages 
because they are identical to those in Table A2.

A few  remarks on the effects of the AC and SC mechanisms are worthy at this point.

The AC term is important in the early contraction pre-main sequence
phase. By comparing AC and Standard models
we notice that, while the absolute duration 
of this phase increases as the mass decreases,
the relative effect (i.e. the ratio between lifetimes at the Hayashi Minimum
for a given mass) increases strongly with the stellar mass.
In any case the AC phase does not particularly affect the 
total  nuclear burning lifetime. Actually some Standard models 
show slightly longer Hydrogen burning lifetimes compared to the corresponding
AC cases. This reflects the fact that the beginning of the main sequence is reached with 
small structural differences  (e.g. location of convective boundaries), 
that propagates during the subsequent evolution.

The SC term is never important in the pre-main sequence contraction
phase since the large stellar radii strongly limit the SC luminosity
(directly via Eq. (\ref{paramLum}) and indirectly through the corrective term due to thermalization time,
Eq. (\ref{tauth})).

\begin{table*} \centering \tiny 
\begin{minipage}{140mm}
\caption{The values for ``standard'', metal--free stars evolving without DM annihilation.}
\begin{tabular}{lrccccccc}
\hline
M/M$_\odot$&Age(Myr)&Log(L/L$_\odot$)&Log(T$_{eff}$)&R$_*$(cm)&Log($\rho_c$)&Log(T$_c$)&X$_c$&L$_{Nuc}$/L$_{*}$\\
\hline \multicolumn{9}{l}{Starting Point}\\
\hline
         5&   0.0000000&     4.762&     3.646&    2.8556 10$^{13}$&    -6.317&     4.898& -& -\\
         7&   0.0000000&     4.854&     3.651&    3.0992 10$^{13}$&    -6.269&     5.008& -& -\\
         9&   0.0000000&     5.096&     3.649&    4.1400 10$^{13}$&    -6.611&     4.962& -& -\\
        12&   0.0000000&     5.223&     3.652&    4.7252 10$^{13}$&    -6.650&     5.014& -& -\\
        15&   0.0000000&     5.305&     3.656&    5.0942 10$^{13}$&    -6.624&     5.068& -& -\\
        20&   0.0000000&     5.416&     3.660&    5.6816 10$^{13}$&    -6.604&     5.129& -& -\\
        40&   0.0000000&     5.769&     3.669&    8.1776 10$^{13}$&    -6.692&     5.211& -& -\\
       100&   0.0000000&     6.178&     3.699&    1.1439 10$^{14}$&    -6.543&     5.377& -& -\\
       200&   0.0000000&     6.471&     3.715&    1.4849 10$^{14}$&    -6.495&     5.466& -& -\\
       400&   0.0000000&     6.782&     3.736&    1.9334 10$^{14}$&    -6.449&     5.547& -& -\\
       600&   0.0000000&     7.001&     3.749&    2.3371 10$^{14}$&    -6.471&     5.575& -& -\\
\hline \multicolumn{9}{l}{Hayashi Minimum Luminosity Model}\\
\hline
         5&   0.0180837&     2.643&     3.739&    1.6218 10$^{12}$&    -2.038&     6.237& -& -\\
         7&   0.0069470&     3.103&     3.730&    2.8721 10$^{12}$&    -2.640&     6.129& -& -\\
         9&   0.0035714&     3.432&     3.724&    4.3114 10$^{12}$&    -3.058&     6.058& -& -\\
        12&   0.0016542&     3.790&     3.717&    6.7326 10$^{12}$&    -3.558&     5.971& -& -\\
        15&   0.0010506&     4.057&     3.713&    9.3245 10$^{12}$&    -3.838&     5.931& -& -\\
        20&   0.0005610&     4.375&     3.708&    1.3782 10$^{13}$&    -4.261&     5.861& -& -\\
        40&   0.0002053&     5.034&     3.703&    3.0049 10$^{13}$&    -4.917&     5.776& -& -\\
       100&   0.0000618&     5.742&     3.707&    6.6582 10$^{13}$&    -5.633&     5.672& -& -\\
       200&   0.0000261&     6.265&     3.726&    1.1135 10$^{14}$&    -5.990&     5.632& -& -\\
       400&   0.0000132&     6.748&     3.754&    1.7054 10$^{14}$&    -6.154&     5.644& -& -\\
       600&   0.0000039&     6.997&     3.757&    2.2503 10$^{14}$&    -6.372&     5.608& -& -\\
\hline \multicolumn{9}{l}{L$_{Nuc.}$/L$_{*}$= 95\%}\\
\hline
         5&   0.7510101&     2.925&     4.453&    8.3722 10$^{10}$&     2.090&     7.676&        0.75194&    9.5000 10$^{-1}$\\
         7&   0.4728800&     3.378&     4.552&    8.9459 10$^{10}$&     2.114&     7.777&        0.75154&    9.5001 10$^{-1}$\\
         9&   0.3380348&     3.705&     4.624&    9.3718 10$^{10}$&     2.139&     7.851&        0.75111&    9.5000 10$^{-1}$\\
        12&   0.2012758&     4.072&     4.703&    9.9462 10$^{10}$&     2.169&     7.931&        0.75197&    9.5000 10$^{-1}$\\
        15&   0.1537508&     4.340&     4.762&    1.0292 10$^{11}$&     2.200&     7.992&        0.75210&    9.5000 10$^{-1}$\\
        20&   0.1005507&     4.699&     4.829&    1.1443 10$^{11}$&     2.201&     8.051&        0.75305&    9.5000 10$^{-1}$\\
        40&   0.0407115&     5.456&     4.921&    1.7910 10$^{11}$&     1.998&     8.100&        0.75440&    9.5001 10$^{-1}$\\
       100&   0.0220905&     6.161&     4.988&    2.9595 10$^{11}$&     1.748&     8.138&        0.75480&    9.5002 10$^{-1}$\\
       200&   0.0167265&     6.599&     5.018&    4.2737 10$^{11}$&     1.585&     8.159&        0.75480&    9.5224 10$^{-1}$\\
       400&   0.0136819&     6.995&     5.033&    6.2788 10$^{11}$&     1.443&     8.177&        0.75490&    9.5000 10$^{-1}$\\
       600&   0.0130140&     7.208&     5.039&    7.8188 10$^{11}$&     1.359&     8.185&        0.75490&    9.5001 10$^{-1}$\\
\hline \multicolumn{9}{l}{Central Hydrogen= 0}\\
\hline
         5&  55.5662916&     3.336&     4.504&    1.0655 10$^{11}$&     3.214&     7.974&        0.00000&    0.0000\\
         7&  29.3981445&     3.728&     4.567&    1.2473 10$^{11}$&     2.916&     8.039&        0.00000&    0.0000\\
         9&  19.8050932&     4.061&     4.602&    1.5618 10$^{11}$&     2.758&     8.077&        0.00000&    0.0000\\
        12&  14.0895636&     4.456&     4.638&    2.0846 10$^{11}$&     2.604&     8.110&        0.00000&    0.0000\\
        15&  10.5241264&     4.716&     4.666&    2.4681 10$^{11}$&     2.520&     8.129&        0.00000&    0.0000\\
        20&   7.6663778&     5.034&     4.695&    3.1161 10$^{11}$&     2.411&     8.150&        0.00000&    0.0000\\
        40&   4.1366357&     5.666&     4.740&    5.2562 10$^{11}$&     2.216&     8.188&        0.00000&    0.0000\\
       100&   2.6233011&     6.311&     4.748&    1.0603 10$^{12}$&     1.981&     8.219&        0.00000&    0.0000\\
       200&   2.1888969&     6.718&     4.736&    1.7944 10$^{12}$&     1.824&     8.236&        0.00000&    0.0000\\
       400&   1.9639194&     7.088&     4.659&    3.9227 10$^{12}$&     1.693&     8.254&        0.00000&    0.0000\\
       600&   2.0630663&     7.418&     4.785&    3.2025 10$^{12}$&     1.594&     8.268&        0.00000&    0.0000\\
\hline 
\end{tabular}
\end{minipage}
\label{standard}
\end{table*}

\begin{table*} \centering \tiny 
\begin{minipage}{140mm}
\caption{AC mechanism only}
\begin{tabular}{lrccccccc}
\hline
M/M$_\odot$&Age(Myr)&Log(L/L$_\odot$)&Log(T$_{eff}$)&R$_*$(cm)&Log($\rho_c$)&Log(T$_c$)&X$_c$&L$_{Nuc}$/L$_{*}$\\
\hline \multicolumn{9}{l}{Starting Point}\\
\hline
         5&   0.0000000&     5.022&     3.637&    4.0174 10$^{13}$&    -6.924&     4.712& -& -\\
         7&   0.0000000&     5.151&     3.642&    4.5638 10$^{13}$&    -6.930&     4.798& -& -\\
         9&   0.0000000&     5.247&     3.645&    5.0209 10$^{13}$&    -6.943&     4.860& -& -\\
        12&   0.0000000&     5.357&     3.648&    5.6097 10$^{13}$&    -6.945&     4.922& -& -\\
        15&   0.0000000&     5.444&     3.651&    6.1288 10$^{13}$&    -6.941&     4.968& -& -\\
        20&   0.0000000&     5.556&     3.655&    6.8376 10$^{13}$&    -6.919&     5.027& -& -\\
        40&   0.0000000&     5.837&     3.667&    8.9564 10$^{13}$&    -6.845&     5.161& -& -\\
       100&   0.0000000&     6.257&     3.699&    1.2491 10$^{14}$&    -6.670&     5.334& -& -\\
       200&   0.0000000&     6.547&     3.715&    1.6202 10$^{14}$&    -6.622&     5.424& -& -\\
       400&   0.0000000&     6.844&     3.733&    2.0996 10$^{14}$&    -6.576&     5.504& -& -\\
       600&   0.0000000&     7.021&     3.747&    2.4173 10$^{14}$&    -6.533&     5.555& -& -\\
\hline \multicolumn{9}{l}{L$_{AC}$/L$_{*}$=50\%}\\
\hline
         5&   0.0096533&     3.074&     3.717&    2.9553 10$^{12}$&    -3.216&     5.937& -& -\\
         7&   0.0151310&     3.201&     3.722&    3.3305 10$^{12}$&    -3.086&     6.018& -& -\\
         9&   0.0177185&     3.449&     3.724&    4.3954 10$^{12}$&    -3.043&     6.067& -& -\\
        12&   0.0175911&     3.886&     3.726&    7.2124 10$^{12}$&    -3.174&     6.095& -& -\\
        15&   0.0163967&     4.187&     3.728&    1.0079 10$^{13}$&    -3.375&     6.086& -& -\\
        20&   0.0144465&     4.546&     3.748&    1.3908 10$^{13}$&    -3.608&     6.081& -& -\\
        40&   0.0092501&     5.259&     3.766&    2.9110 10$^{13}$&    -4.295&     5.992& -& -\\
       100&   0.0053198&     6.040&     3.943&    3.1780 10$^{13}$&    -4.055&     6.204& -& -\\
       200&   0.0036779&     6.514&     4.041&    3.4862 10$^{13}$&    -3.914&     6.326& -& -\\
       400&   0.0024905&     6.935&     4.094&    4.4213 10$^{13}$&    -3.871&     6.406& -& -\\
       600&   0.0020670&     7.164&     4.110&    5.3619 10$^{13}$&    -3.881&     6.439& -& -\\
\hline \multicolumn{9}{l}{L$_{AC}$/L$_{*}$=10\%}\\
\hline
         5&   0.0518843&     2.801&     3.763&    1.7432 10$^{12}$&    -1.456&     6.419& -& -\\
         7&   0.0419409&     3.354&     3.891&    1.8255 10$^{12}$&    -1.346&     6.559& -& -\\
         9&   0.0341297&     3.719&     3.969&    1.9446 10$^{12}$&    -1.371&     6.623& -& -\\
        12&   0.0287672&     4.110&     4.070&    1.9156 10$^{12}$&    -1.305&     6.726& -& -\\
        15&   0.0250591&     4.388&     4.134&    1.9676 10$^{12}$&    -1.293&     6.788& -& -\\
        20&   0.0215014&     4.719&     4.221&    1.9227 10$^{12}$&    -1.205&     6.888& -& -\\
        40&   0.0136990&     5.392&     4.365&    2.1521 10$^{12}$&    -1.184&     7.036& -& -\\
       100&   0.0076639&     6.099&     4.454&    3.2231 10$^{12}$&    -1.376&     7.098& -& -\\
       200&   0.0057030&     6.553&     4.527&    3.8831 10$^{12}$&    -1.310&     7.194& -& -\\
       400&   0.0042986&     6.959&     4.571&    5.0526 10$^{12}$&    -1.316&     7.258& -& -\\
       600&   0.0037723&     7.180&     4.586&    6.0964 10$^{12}$&    -1.352&     7.281& -& -\\
\hline \multicolumn{9}{l}{Hayashi Minimum Luminosity Model}\\
\hline
         5&   0.0318454&     2.650&     3.739&    1.6363 10$^{12}$&    -2.062&     6.232& -& -\\
         7&   0.0207489&     3.115&     3.730&    2.9142 10$^{12}$&    -2.650&     6.130& -& -\\
         9&   0.0177888&     3.449&     3.724&    4.3936 10$^{12}$&    -3.036&     6.069& -& -\\
        12&   0.0153947&     3.814&     3.717&    6.9172 10$^{12}$&    -3.536&     5.982& -& -\\
        15&   0.0144330&     4.089&     3.713&    9.6710 10$^{12}$&    -3.842&     5.936& -& -\\
        20&   0.0124400&     4.410&     3.707&    1.4378 10$^{13}$&    -4.283&     5.859& -& -\\
        40&   0.0082365&     5.094&     3.699&    3.2837 10$^{13}$&    -5.150&     5.707& -& -\\
       100&   0.0043604&     5.805&     3.704&    7.2797 10$^{13}$&    -5.820&     5.613& -& -\\
       200&   0.0024627&     6.297&     3.721&    1.1845 10$^{14}$&    -6.130&     5.587& -& -\\
       400&   0.0009528&     6.748&     3.739&    1.8375 10$^{14}$&    -6.366&     5.574& -& -\\
       600&   0.0004468&     6.998&     3.756&    2.2582 10$^{14}$&    -6.386&     5.603& -& -\\
\hline \multicolumn{9}{l}{L$_{Nuc.}$/L$_{*}$= 95\%}\\
\hline
         5&   0.6821600&     2.928&     4.452&    8.4617 10$^{10}$&     2.079&     7.672&        0.75305&    9.5000 10$^{-1}$\\
         7&   0.4372555&     3.380&     4.551&    9.0212 10$^{10}$&     2.105&     7.774&        0.75265&    9.5000 10$^{-1}$\\
         9&   0.3242375&     3.707&     4.623&    9.4319 10$^{10}$&     2.133&     7.849&        0.75211&    9.5000 10$^{-1}$\\
        12&   0.2193152&     4.071&     4.703&    9.9367 10$^{10}$&     2.169&     7.931&        0.75208&    9.5000 10$^{-1}$\\
        15&   0.1737158&     4.342&     4.762&    1.0320 10$^{11}$&     2.198&     7.992&        0.75206&    9.5000 10$^{-1}$\\
        20&   0.1152623&     4.699&     4.829&    1.1440 10$^{11}$&     2.201&     8.051&        0.75310&    9.5000 10$^{-1}$\\
        40&   0.0510427&     5.455&     4.921&    1.7890 10$^{11}$&     1.999&     8.100&        0.75440&    9.5001 10$^{-1}$\\
       100&   0.0278809&     6.161&     4.988&    2.9578 10$^{11}$&     1.749&     8.138&        0.75480&    9.5001 10$^{-1}$\\
       200&   0.0208053&     6.599&     5.018&    4.2722 10$^{11}$&     1.586&     8.159&        0.75480&    9.5001 10$^{-1}$\\
       400&   0.0168216&     6.995&     5.033&    6.2819 10$^{11}$&     1.442&     8.177&        0.75490&    9.5001 10$^{-1}$\\
       600&   0.0152766&     7.208&     5.039&    7.8170 10$^{11}$&     1.360&     8.185&        0.75490&    9.5080 10$^{-1}$\\
\hline \multicolumn{9}{l}{Central Hydrogen= 0 or Scattering Stallation model}\\
\hline
         5&  56.1465439&     3.340&     4.505&    1.0660 10$^{11}$&     3.196&     7.983&        0.00000&    0.0000\\
         7&  29.1142533&     3.725&     4.569&    1.2323 10$^{11}$&     2.929&     8.037&        0.00000&    0.0000\\
         9&  19.2892971&     4.046&     4.607&    1.4988 10$^{11}$&     2.772&     8.074&        0.00000&    0.0000\\
        12&  13.2154438&     4.420&     4.646&    1.9311 10$^{11}$&     2.624&     8.107&        0.00000&    0.0000\\
        15&  10.7956074&     4.734&     4.663&    2.5626 10$^{11}$&     2.515&     8.131&        0.00000&    0.0000\\
        20&   7.6619203&     5.035&     4.695&    3.1251 10$^{11}$&     2.415&     8.151&        0.00000&    0.0000\\
        40&   4.1429360&     5.667&     4.740&    5.2378 10$^{11}$&     2.219&     8.189&        0.00000&    0.0000\\
       100&   2.6907600&     6.325&     4.740&    1.1194 10$^{12}$&     1.986&     8.222&        0.00000&    0.0000\\
       200&   2.1910250&     6.720&     4.712&    2.0064 10$^{12}$&     1.831&     8.239&        0.00000&    0.0000\\
       400&   2.2893942&     7.330&     5.117&    6.2717 10$^{11}$&     1.663&     8.269&        0.00000&    0.0000\\
       600&   1.8355234&     7.289&     4.543&    8.4022 10$^{12}$&     1.612&     8.261&        0.00000&    0.0000\\
\hline 
\end{tabular}
\end{minipage}
\label{ac}
\end{table*}

\begin{table*} \centering \tiny 
\begin{minipage}{140mm}
\caption{AC and SC mechanisms active; $\rho_\chi$= 10$^{11}$GeV/cm$^3$,  $\sigma_0$= 10$^{-39}$cm$^2$  }
\begin{tabular}{lrccccccc}
\hline
M/M$_\odot$&Age(Myr)&Log(L/L$_\odot$)&Log(T$_{eff}$)&R$_*$(cm)&Log($\rho_c$)&Log(T$_c$)&X$_c$&L$_{Nuc}$/L$_{*}$\\
\hline \multicolumn{9}{l}{Hayashi Minimum Luminosity Model}\\
\hline
         5&   0.0321649&     2.650&     3.739&    1.6353 10$^{12}$&    -2.056&     6.233& -& -\\
         7&   0.0207489&     3.115&     3.730&    2.9142 10$^{12}$&    -2.650&     6.130& -& -\\
         9&   0.0177888&     3.449&     3.724&    4.3936 10$^{12}$&    -3.036&     6.069& -& -\\
        12&   0.0153947&     3.814&     3.717&    6.9172 10$^{12}$&    -3.536&     5.982& -& -\\
        15&   0.0144330&     4.089&     3.713&    9.6710 10$^{12}$&    -3.842&     5.936& -& -\\
        20&   0.0124400&     4.410&     3.707&    1.4378 10$^{13}$&    -4.283&     5.859& -& -\\
        40&   0.0082365&     5.094&     3.699&    3.2837 10$^{13}$&    -5.150&     5.707& -& -\\
       100&   0.0043604&     5.805&     3.704&    7.2797 10$^{13}$&    -5.820&     5.613& -& -\\
       200&   0.0024627&     6.297&     3.721&    1.1845 10$^{14}$&    -6.130&     5.587& -& -\\
       300&   0.0016023&     6.565&     3.732&    1.5320 10$^{14}$&    -6.254&     5.585& -& -\\
       400&   0.0009528&     6.748&     3.739&    1.8375 10$^{14}$&    -6.366&     5.574& -& -\\
       600&   0.0004468&     6.998&     3.756&    2.2582 10$^{14}$&    -6.386&     5.603& -& -\\
\hline \multicolumn{9}{l}{L$_{Nuc.}$/L$_{*}$= 95\%}\\
\hline
         7&  58.2469694&     3.692&     4.559&    1.2448 10$^{11}$&     2.532&     7.974&        0.11475&    9.5000 10$^{-1}$\\
         9&  30.9662677&     3.992&     4.607&    1.4085 10$^{11}$&     2.378&     7.985&        0.15375&    9.5000 10$^{-1}$\\
        12&  15.8442989&     4.330&     4.663&    1.6080 10$^{11}$&     2.226&     7.997&        0.21329&    9.5000 10$^{-1}$\\
        15&  10.3999383&     4.586&     4.707&    1.7632 10$^{11}$&     2.117&     8.006&        0.27733&    9.5000 10$^{-1}$\\
        20&   6.3947664&     4.879&     4.758&    1.9554 10$^{11}$&     2.001&     8.018&        0.34013&    9.5000 10$^{-1}$\\
        40&   2.4225710&     5.507&     4.856&    2.5643 10$^{11}$&     1.767&     8.043&        0.46558&    9.5000 10$^{-1}$\\
       100&   1.1063569&     6.203&     4.935&    3.9656 10$^{11}$&     1.529&     8.075&        0.57258&    9.5179 10$^{-1}$\\
       200&   0.9510749&     6.633&     4.950&    6.0802 10$^{11}$&     1.356&     8.089&        0.53188&    9.4999 10$^{-1}$\\
       300&   0.9983394&     6.866&     4.946&    8.1006 10$^{11}$&     1.263&     8.096&        0.48495&    9.4997 10$^{-1}$\\
       400&   1.1107156&     7.027&     4.933&    1.0328 10$^{12}$&     1.200&     8.101&        0.42339&    9.5000 10$^{-1}$\\
       600&   0.9687933&     7.246&     4.963&    1.1603 10$^{12}$&     1.121&     8.109&        0.45154&    9.4998 10$^{-1}$\\
\hline \multicolumn{9}{l}{Central Hydrogen= 0 or Scattering Stallation model}\\
\hline
         5 $^s$&   5.0000000&     2.830&     4.322&    1.3769 10$^{11}$&     1.384&     7.452&        0.75400&    4.8940 10$^{-2}$\\
         7&  60.2205803&     3.758&     4.563&    1.3173 10$^{11}$&     2.903&     8.048&        0.00000&    0.0000\\
         9&  32.9417891&     4.076&     4.603&    1.5812 10$^{11}$&     2.758&     8.079&        0.00000&    0.0000\\
        12&  17.9675788&     4.441&     4.642&    2.0086 10$^{11}$&     2.615&     8.109&        0.00000&    0.0000\\
        15&  12.6592264&     4.728&     4.665&    2.5143 10$^{11}$&     2.517&     8.131&        0.00000&    0.0000\\
        20&   8.6058962&     5.038&     4.695&    3.1378 10$^{11}$&     2.414&     8.151&        0.00000&    0.0000\\
        40&   4.3902484&     5.670&     4.740&    5.2772 10$^{11}$&     2.218&     8.190&        0.00000&    0.0000\\
       100&   2.9876710&     6.358&     4.609&    2.1264 10$^{12}$&     1.978&     8.225&        0.00000&    0.0000\\
       200&   2.3571500&     6.730&     4.639&    2.8487 10$^{12}$&     1.834&     8.240&        0.00000&    0.0000\\
       300&   2.1357779&     6.942&     4.651&    3.4292 10$^{12}$&     1.747&     8.248&        0.00000&    0.0000\\
       400&   2.0465587&     7.106&     4.637&    4.4207 10$^{12}$&     1.691&     8.255&        0.00000&    0.0000\\
       600&   2.0407706&     7.362&     4.810&    2.6816 10$^{12}$&     1.595&     8.265&        0.00000&    0.0000\\
\hline 
\multicolumn{5}{l}{$^s$ Track Stalled}\\
\end{tabular}
\end{minipage}
\label{acsc39}
\end{table*}

\begin{table*} \centering \tiny 
\begin{minipage}{140mm}
\caption{AC and SC mechanisms active; $\rho_\chi$= 10$^{11}$GeV/cm$^3$, $\sigma_0$= 10$^{-38}$cm$^2$  }
\begin{tabular}{lrccccccc}
\hline
M/M$_\odot$&Age(Myr)&Log(L/L$_\odot$)&Log(T$_{eff}$)&R$_*$(cm)&Log($\rho_c$)&Log(T$_c$)&X$_c$&L$_{Nuc}$/L$_{*}$\\
\hline \multicolumn{9}{l}{Hayashi Minimum Luminosity Model}\\
\hline
         5&   0.0732837&     2.865&     3.859&    1.2047 10$^{12}$&    -0.975&     6.618& -& -\\
         7&   0.0207489&     3.115&     3.730&    2.9142 10$^{12}$&    -2.650&     6.130& -& -\\
         9&   0.0177888&     3.449&     3.724&    4.3936 10$^{12}$&    -3.036&     6.069& -& -\\
        12&   0.0153947&     3.814&     3.717&    6.9172 10$^{12}$&    -3.536&     5.982& -& -\\
        15&   0.0144330&     4.089&     3.713&    9.6710 10$^{12}$&    -3.842&     5.936& -& -\\
        20&   0.0124400&     4.410&     3.707&    1.4378 10$^{13}$&    -4.283&     5.859& -& -\\
        40&   0.0082365&     5.094&     3.699&    3.2837 10$^{13}$&    -5.150&     5.707& -& -\\
       100&   0.0043604&     5.805&     3.704&    7.2797 10$^{13}$&    -5.820&     5.613& -& -\\
       200&   0.0030809&     6.284&     3.723&    1.1551 10$^{14}$&    -6.077&     5.605& -& -\\
       400&   0.0016042&     6.745&     3.748&    1.7519 10$^{14}$&    -6.234&     5.618& -& -\\
       600&   0.0004468&     6.998&     3.756&    2.2582 10$^{14}$&    -6.386&     5.603& -& -\\
\hline \multicolumn{9}{l}{L$_{Nuc.}$/L$_{*}$= 95\%}\\
\hline
        40&  20.7324568&     5.628&     4.754&    4.7132 10$^{11}$&     1.800&     8.058&        0.10605&    9.5000 10$^{-1}$\\
       100&   4.5980377&     6.279&     4.792&    8.3431 10$^{11}$&     1.540&     8.080&        0.12656&    9.5133 10$^{-1}$\\
       200&   3.5912405&     6.718&     4.748&    1.7011 10$^{12}$&     1.375&     8.094&        0.14434&    9.5000 10$^{-1}$\\
       400&   3.4325657&     7.085&     4.740&    2.6877 10$^{12}$&     1.241&     8.110&        0.11646&    9.4998 10$^{-1}$\\
       600&   3.9629294&     7.308&     4.652&    5.2214 10$^{12}$&     1.185&     8.122&        0.08405&    9.5000 10$^{-1}$\\
\hline \multicolumn{9}{l}{Central Hydrogen= 0 or Scattering Stallation model}\\
\hline
         5 $^s$&   5.0000000&     2.698&     3.732&    1.7904 10$^{12}$&    -2.567&     6.154&        0.75500&    0.0000\\
         7 $^s$&   5.0000000&     3.191&     3.956&    1.1241 10$^{12}$&    -1.067&     6.721&        0.75500&    1.5187 10$^{-8}$\\
         9 $^s$&   5.0000000&     3.578&     4.119&    8.2850 10$^{11}$&    -0.652&     6.925&        0.75500&    3.4554 10$^{-7}$\\
        12 $^s$&   5.0000000&     3.984&     4.309&    5.5088 10$^{11}$&    -0.068&     7.190&        0.75500&    1.6556 10$^{-5}$\\
        15 $^s$&   5.0000000&     4.274&     4.438&    4.2336 10$^{11}$&     0.342&     7.376&        0.75500&    2.6014 10$^{-4}$\\
        20 $^s$&   5.0000000&     4.630&     4.581&    3.3130 10$^{11}$&     0.774&     7.578&        0.75440&    2.2035 10$^{-3}$\\
        40&  21.0633259&     5.674&     4.735&    5.4184 10$^{11}$&     2.218&     8.190&        0.00000&    0.0000\\
       100&   4.8774489&     6.311&     4.749&    1.0565 10$^{12}$&     1.984&     8.220&        0.00000&    0.0000\\
       200&   3.8658341&     6.744&     4.619&    3.1638 10$^{12}$&     1.833&     8.241&        0.00000&    0.0000\\
       400&   3.6641729&     7.105&     4.454&    1.0244 10$^{13}$&     1.702&     8.256&        0.00000&    0.0000\\
       600&   4.1027411&     7.322&     4.418&    1.5577 10$^{13}$&     1.626&     8.265&        0.00000&    0.0000\\
\hline 
\multicolumn{5}{l}{$^s$ Track Stalled}\\
\end{tabular}
\end{minipage}
\label{acsc38}
\end{table*}

\begin{table*} \centering \tiny 
\begin{minipage}{140mm}
\caption{AC mechanism only, for a neutralino mass $m_\chi$=200GeV.}
\begin{tabular}{lrccccccc}
\hline
M/M$_\odot$&Age(Myr)&Log(L/L$_\odot$)&Log(T$_{eff}$)&R$_*$(cm)&Log($\rho_c$)&Log(T$_c$)&X$_c$&L$_{Nuc}$/L$_{*}$\\
\hline \multicolumn{9}{l}{Starting Point}\\
\hline
         5&   0.0000000&     4.882&     3.642&    3.3438 10$^{13}$&    -6.576&     4.810& -& -\\
         7&   0.0000000&     5.009&     3.647&    3.7899 10$^{13}$&    -6.601&     4.901& -& -\\
         9&   0.0000000&     5.104&     3.649&    4.1733 10$^{13}$&    -6.615&     4.960& -& -\\
        12&   0.0000000&     5.212&     3.653&    4.6489 10$^{13}$&    -6.615&     5.025& -& -\\
        15&   0.0000000&     5.236&     3.659&    4.6423 10$^{13}$&    -6.465&     5.119& -& -\\
        20&   0.0000000&     5.417&     3.660&    5.6860 10$^{13}$&    -6.603&     5.129& -& -\\
        40&   0.0000000&     5.699&     3.673&    7.4259 10$^{13}$&    -6.532&     5.264& -& -\\
       100&   0.0000000&     6.099&     3.698&    1.0491 10$^{14}$&    -6.421&     5.417& -& -\\
       200&   0.0000000&     6.400&     3.716&    1.3640 10$^{14}$&    -6.370&     5.508& -& -\\
       400&   0.0000000&     6.763&     3.752&    1.7578 10$^{14}$&    -6.187&     5.633& -& -\\
       600&   0.0000000&     7.030&     3.786&    2.0377 10$^{14}$&    -6.004&     5.730& -& -\\
\hline \multicolumn{9}{l}{L$_{AC}$/L$_{*.}$=50\%}\\
\hline
         5&   0.0080226&     3.090&     3.716&    3.0198 10$^{12}$&    -3.244&     5.928& -& -\\
         7&   0.0123634&     3.216&     3.722&    3.3996 10$^{12}$&    -3.128&     6.007& -& -\\
         9&   0.0139301&     3.452&     3.723&    4.4418 10$^{12}$&    -3.139&     6.040& -& -\\
        12&   0.0140096&     3.872&     3.724&    7.1618 10$^{12}$&    -3.235&     6.074& -& -\\
        15&   0.0116689&     4.174&     3.724&    1.0123 10$^{13}$&    -3.451&     6.060& -& -\\
        20&   0.0108273&     4.537&     3.738&    1.4423 10$^{13}$&    -3.669&     6.059& -& -\\
        40&   0.0070071&     5.266&     3.778&    2.7769 10$^{13}$&    -4.223&     6.017& -& -\\
       100&   0.0039907&     6.043&     3.975&    2.7385 10$^{13}$&    -3.902&     6.255& -& -\\
       200&   0.0025829&     6.516&     4.059&    3.2085 10$^{13}$&    -3.826&     6.355& -& -\\
       400&   0.0016259&     6.935&     4.111&    4.0936 10$^{13}$&    -3.796&     6.431& -& -\\
       600&   0.0013353&     7.164&     4.129&    4.9076 10$^{13}$&    -3.795&     6.467& -& -\\
\hline \multicolumn{9}{l}{L$_{AC}$/L$_{*}$=10\%}\\
\hline
         5&   0.0509513&     2.806&     3.765&    1.7354 10$^{12}$&    -1.431&     6.428& -& -\\
         7&   0.0406938&     3.361&     3.907&    1.7131 10$^{12}$&    -1.271&     6.585& -& -\\
         9&   0.0315087&     3.723&     3.980&    1.8568 10$^{12}$&    -1.318&     6.642& -& -\\
        12&   0.0257831&     4.113&     4.081&    1.8228 10$^{12}$&    -1.248&     6.745& -& -\\
        15&   0.0207747&     4.390&     4.144&    1.8789 10$^{12}$&    -1.240&     6.806& -& -\\
        20&   0.0182315&     4.720&     4.235&    1.8072 10$^{12}$&    -1.131&     6.912& -& -\\
        40&   0.0115195&     5.392&     4.374&    2.0643 10$^{12}$&    -1.133&     7.053& -& -\\
       100&   0.0063115&     6.100&     4.458&    3.1661 10$^{12}$&    -1.354&     7.105& -& -\\
       200&   0.0044895&     6.553&     4.521&    3.9951 10$^{12}$&    -1.346&     7.182& -& -\\
       400&   0.0032827&     6.958&     4.559&    5.3438 10$^{12}$&    -1.387&     7.234& -& -\\
       600&   0.0029023&     7.180&     4.574&    6.4304 10$^{12}$&    -1.418&     7.259& -& -\\
\hline \multicolumn{9}{l}{Hayashi Minimum Luminosity Model}\\
\hline
         5&   0.0300699&     2.650&     3.739&    1.6362 10$^{12}$&    -2.059&     6.233& -& -\\
         7&   0.0182169&     3.115&     3.730&    2.9145 10$^{12}$&    -2.650&     6.130& -& -\\
         9&   0.0148365&     3.449&     3.724&    4.3933 10$^{12}$&    -3.037&     6.068& -& -\\
        12&   0.0122626&     3.814&     3.717&    6.9151 10$^{12}$&    -3.539&     5.981& -& -\\
        15&   0.0101071&     4.088&     3.713&    9.6724 10$^{12}$&    -3.855&     5.932& -& -\\
        20&   0.0091717&     4.409&     3.707&    1.4361 10$^{13}$&    -4.275&     5.861& -& -\\
        40&   0.0061004&     5.091&     3.700&    3.2531 10$^{13}$&    -5.074&     5.731& -& -\\
       100&   0.0026849&     5.807&     3.703&    7.3077 10$^{13}$&    -5.838&     5.608& -& -\\
       200&   0.0009983&     6.297&     3.721&    1.1832 10$^{14}$&    -6.124&     5.589& -& -\\
       400&   0.0000092&     6.761&     3.752&    1.7546 10$^{14}$&    -6.186&     5.634& -& -\\
       600&   0.0000092&     7.029&     3.786&    2.0349 10$^{14}$&    -6.002&     5.730& -& -\\
\hline \multicolumn{9}{l}{L$_{Nuc.}$/L$_{*}$= 95\%}\\
\hline
         5&   0.6992481&     2.927&     4.452&    8.4513 10$^{10}$&     2.080&     7.673&        0.75283&    9.5000 10$^{-1}$\\
         7&   0.4522332&     3.380&     4.551&    8.9905 10$^{10}$&     2.108&     7.775&        0.75226&    9.5000 10$^{-1}$\\
         9&   0.3378005&     3.705&     4.623&    9.3935 10$^{10}$&     2.137&     7.850&        0.75159&    9.5000 10$^{-1}$\\
        12&   0.2151746&     4.071&     4.703&    9.9326 10$^{10}$&     2.170&     7.931&        0.75200&    9.5000 10$^{-1}$\\
        15&   0.1683067&     4.342&     4.762&    1.0317 10$^{11}$&     2.199&     7.992&        0.75207&    9.5000 10$^{-1}$\\
        20&   0.1110554&     4.699&     4.829&    1.1438 10$^{11}$&     2.201&     8.051&        0.75310&    9.5000 10$^{-1}$\\
        40&   0.0483624&     5.455&     4.921&    1.7885 10$^{11}$&     1.999&     8.100&        0.75440&    9.5000 10$^{-1}$\\
       100&   0.0261951&     6.161&     4.988&    2.9566 10$^{11}$&     1.749&     8.139&        0.75480&    9.5002 10$^{-1}$\\
       200&   0.0193638&     6.599&     5.018&    4.2712 10$^{11}$&     1.586&     8.159&        0.75480&    9.5003 10$^{-1}$\\
       400&   0.0156433&     6.994&     5.033&    6.2776 10$^{11}$&     1.443&     8.177&        0.75490&    9.6414 10$^{-1}$\\
       600&   0.0142502&     7.208&     5.039&    7.8114 10$^{11}$&     1.361&     8.186&        0.75490&    9.6439 10$^{-1}$\\
\hline \multicolumn{9}{l}{Central Hydrogen= 0 or Scattering Stallation model}\\
\hline
         5&  56.1384003&     3.340&     4.505&    1.0644 10$^{11}$&     3.200&     7.983&        0.00000&    0.0000\\
         7&  29.7313755&     3.734&     4.566&    1.2641 10$^{11}$&     2.924&     8.040&        0.00000&    0.0000\\
         9&  19.5580299&     4.053&     4.604&    1.5289 10$^{11}$&     2.771&     8.076&        0.00000&    0.0000\\
        12&  13.7294256&     4.447&     4.640&    2.0425 10$^{11}$&     2.613&     8.110&        0.00000&    0.0000\\
        15&  10.7102756&     4.726&     4.665&    2.5130 10$^{11}$&     2.518&     8.131&        0.00000&    0.0000\\
        20&   7.6979362&     5.040&     4.694&    3.1552 10$^{11}$&     2.415&     8.152&        0.00000&    0.0000\\
        40&   4.1505195&     5.668&     4.741&    5.2413 10$^{11}$&     2.222&     8.190&        0.00000&    0.0000\\
       100&   2.7010537&     6.327&     4.735&    1.1476 10$^{12}$&     1.982&     8.221&        0.00000&    0.0000\\
       200&   2.2941418&     6.744&     4.547&    4.4148 10$^{12}$&     1.843&     8.242&        0.00000&    0.0000\\
       400 $^a$&   1.9648883&     7.194&     4.342&    1.9067 10$^{13}$&     1.191&     8.105&        0.16920&    9.7490 10$^{-1}$\\
       600&   2.0679154&     7.538&     5.111&    8.2005 10$^{11}$&     1.583&     8.275&        0.00000&    0.0000\\
\hline  
\multicolumn{5}{l}{$^a$  Problems to reach end of H-burning}\\
\end{tabular}
\end{minipage}
\label{dm200}
\end{table*}

\bsp
\label{lastpage}
\end{document}